\documentclass[12pt]{article}
\usepackage{amsmath,amsfonts, amssymb}
\usepackage{hyperref}
\usepackage{graphicx}
\usepackage{psfrag}

\newcommand{\bea}{\begin{eqnarray}}
\newcommand{\eea}{\end{eqnarray}}
\newcommand{\ra}{\rightarrow}

\newcommand{\be}{\begin{equation}}
\newcommand{\ee}{\end{equation}}
\newcommand{\ba}{\begin{eqnarray}}
\newcommand{\ea}{\end{eqnarray}}
\newcommand{\bi}{\begin{itemize}}
\newcommand{\ei}{\end{itemize}}
\newcommand{\tr}{{\rm tr}}
\newcommand{\Tr}{{\rm Tr}}

\newcommand{\R}{\mathbb{R}}
\newcommand{\C}{{\mathbb C}}
\newcommand{\p}{\partial}

\newcommand{\N}{{\mathcal N}}

\newcommand{\Ncal}{{\mathcal N}}

\newcommand{\Ocal}{{\mathcal O}}

\newcommand{\Lcal}{{\mathcal L}}
\newcommand{\Wcal}{{\mathcal W}}

\newcommand{\Rcal}{{\mathcal R}}

\newcommand{\gfrak}{{\mathfrak g}}

\newcommand{\tfrak}{{\mathfrak t}}

\newcommand{\nn}{\nonumber}
\newcommand{\mo}{{-1}} 

\newcommand{\f}{\frac}
\newcommand{\half}{\frac{1}{2}}
\newcommand{\oo}{\frac{1}}

\newcommand{\aslash}[1]{\,\,{\raise.15ex\hbox{/}\mkern-12mu #1}}
\newcommand{\bslash}[1]{\,\,{\raise.15ex\hbox{/}\mkern-9mu #1}}

\newcommand{\LR}{{}^L\negthinspace R}
\newcommand{\LG}{{}^L\negthinspace\hspace{.4mm} G}

\newcommand{\Lw}{{}^L\negthinspace\hspace{.4mm} w}

\newcommand{\Lalpha}{{}^L\negthinspace\hspace{.4mm} \alpha}

\newcommand{\sfg}{{\mathsf g}}

\newcommand{\LL}{{}^L\negthinspace}
\newcommand{\Lt}{{}^L\negthinspace\hspace{.4mm}  \mathfrak t}

\newcommand{\Ltau}{{}^L\negthinspace\hspace{.4mm}  \tau}
\newcommand{\LLtau}{\raisebox{.8mm}{\mbox{\it \scriptsize{L}}}\mspace{.3mu}\tau}
\newcommand{\Lgc}{{}^L\negthinspace\hspace{.2mm} g}
\newcommand{\LLgc}{\raisebox{.8mm}{\mbox{\it \scriptsize{L}}} g }
\newcommand{\Ltheta}{{}^L\negthinspace\hspace{.4mm} \theta}

\renewcommand{\bar}{\overline}
\renewcommand{\tilde}{\widetilde}
\renewcommand{\hat}{\widehat}

\renewcommand{\title}[1]{\vbox{\center\LARGE{#1}}\vspace{5mm}}
\renewcommand{\author}[1]{\vbox{\center#1}\vspace{5mm}}

\begin{document}
\bibliographystyle{utphys}

\begin{titlepage}

\bigskip\bigskip

\begin{center}
\rightline{ }
\vskip 18mm
 
\centerline{\LARGE 
$S$-duality,  't Hooft operators and }

\medskip\medskip
\centerline{\LARGE
the operator product expansion
}
 
\vskip 15mm
{Jaume Gomis\footnote{\href{mailto: jgomis@perimeterinstitute.ca}{\rm jgomis@perimeterinstitute.ca}}
~~ and~~  Takuya Okuda\footnote{\href{mailto: takuya@perimeterinstitute.ca}{\rm takuya@perimeterinstitute.ca}}}
\vskip 10mm
{\it Perimeter Institute for Theoretical Physics,

Waterloo, Ontario, N2L 2Y5, Canada
}

\vskip 2cm

\end{center}

\abstract{

\noindent \normalsize{ 
\noindent
We study $S$-duality in $\Ncal=4$ super Yang-Mills
with an arbitrary gauge group
by determining the operator product expansion
of the  circular BPS Wilson and 't Hooft loop operators.
The  coefficients in the expansion of an 't Hooft loop operator for chiral primary operators and the stress-energy tensor 
are calculated in perturbation theory using the   quantum path-integral definition of the 't Hooft operator recently proposed.
The corresponding  operator product coefficients for the dual  Wilson loop operator are determined in the
strong coupling expansion. The results for the 't Hooft operator in the weak coupling expansion exactly reproduce those for the dual Wilson loop operator    in the
strong coupling expansion, thereby
demonstrating  the quantitative prediction of $S$-duality for these observables.
 }
 }

\vfill

\end{titlepage}

\tableofcontents


\section{Introduction and summary}
\numberwithin{equation}{section}

Electric-magnetic duality,  such as $S$-duality
in $\Ncal=4$ super Yang-Mills, 
 maps electrically charged excitations in one theory
to magnetically charged ones in the dual theory.
In the conventional formulation of gauge theories, magnetically charged
objects are not included as integration variables in the path integral. Rather,
they are  realized via  non-trivial field configurations of the electric variables.
The magnetic analog of the Wilson loop operator \cite{Wilson:1974sk}, {\it i.e.} the 't Hooft loop operator \cite{'tHooft:1977hy}, which inserts a magnetically charged source,  is defined by a singular   field configuration of the electric variables.

Even though an 't Hooft loop  is a {\it disorder operator},
defined by   prescribing a singularity along the loop,  
it    shares  features   associated with ordinary operators,
which are characterized by gauge invariant functions of the electric
variables of  the theory.
For example, just as the potential generated by a distribution of charges admits
a multi-pole expansion, 
any loop operator -- an     't Hooft ($T$) or Wilson ($W$) operator --  appears as an infinite series of local operators
to an observer who probes the loop operator
from a distance much larger than the size of the loop:
\begin{center}
\psfrag{TorW}{$T$ or $W$}
\psfrag{=}{$=$}
\psfrag{1}{ $\displaystyle \sum_i ~~~~~ \Ocal_i$}
\psfrag{O}{}
\hspace{7mm}\includegraphics[width=100mm]{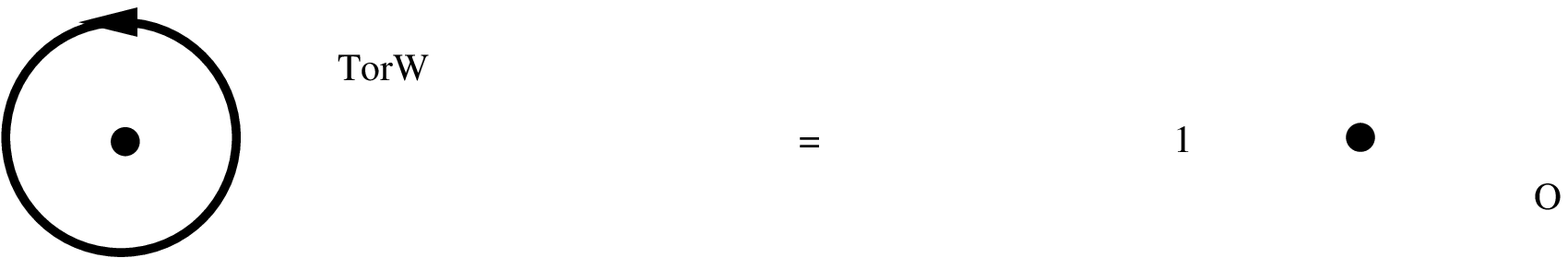}
\end{center}
\begin{center}
{\rm Figure 1. The operator product expansion of a loop operator.}
\end{center}
Therefore  an 't Hooft operator, despite being  a disorder operator,
also admits an operator product expansion (OPE)  in terms of an 
infinite sum of local   operators $\Ocal_i$.

A suitable arena where to study the OPE of loop operators and the action of electric-magnetic duality on the OPE
is $\Ncal=4$ super Yang-Mill theory. $S$-duality \cite{Montonen:1977sn, Witten:1978ma,Osborn:1979tq} posits that ${\cal N}=4$ super Yang-Mills with gauge group $G$ and coupling constant $\tau$ 
is equivalent to   ${\cal N}=4$ super Yang-Mills with dual gauge group $\LG$
\cite{Goddard:1976qe} and dual coupling constant $\LLtau$. The coupling constants of the dual theories are related by the strong-weak coupling  transformation
\ba
\LLtau=-\f{1}{n_\gfrak \tau}\,, \nonumber
\ea
where
\ba 
\tau={\theta\over 2\pi}+{4\pi i\over g^2}\,,
~~~
~~~~\LLtau={\Ltheta\over 2\pi}+{4\pi i\over (\LLgc)^2}\,,
\nonumber
\ea
and where $n_\gfrak=1,2$ or $3$ depending\footnote{\label{ng} Here
$n_\gfrak=1$ for simply laced algebras; $n_\gfrak=2$ for
$\mathfrak{so}(2n+1) ,\mathfrak{sp}(n)$ and $\mathfrak{f}_4$; and
$n_\gfrak=3$ for $\mathfrak{g}_2$.} on the choice of gauge group
$G$.

In ${\cal N}=4$ super Yang-Mills,  't Hooft loop operators     in the theory with gauge group $G$ are conjectured to transform under the action of $S$-duality into Wilson loop operators   in the dual theory, which has    gauge group $\LG$. Under $S$-duality
electric and magnetic sources are  exchanged \cite{Kapustin:2005py}
\ba
T(\LR)\longleftrightarrow W(\LR)\,. \nonumber
\ea
 $\LR$ is an irreducible representation of $\LG$, which labels \cite{Kapustin:2005py}
 an 't Hooft operator $T(\LR)$ in the theory with gauge group $G$,   as well  as a  Wilson operator $W(\LR)$ in the theory with gauge group $\LG$.

The recent paper \cite{Gomis:2009ir} has 
 explicitly demostrated that the prediction of $S$-duality for the observables  
\ba 
\left\langle T(\LR)\right\rangle_{G,\tau}= \left\langle
W(\LR)\right\rangle_{\LG,\Ltau}\, \label{prediccion}
\ea
holds to next to leading order in the coupling constant expansion   for  supersymmetric circular loops. This equality was proven by first giving a quantum  definition of an 't Hooft operator, computing its expectation value  to next to leading order in the weak coupling expansion and comparing the result with the strong coupling expansion of the Wilson loop expectation value in the  dual theory \cite{Gomis:2009ir}.

$S$-duality conjecturally acts on all the gauge invariant operators   in ${\cal N}=4$ super Yang-Mills, both on local and non-local operators, thus defining an  isomorphism between  the operators in the two dual descriptions 
\ba {\cal O}\longleftrightarrow {}^L\negthinspace\hspace{.4mm} {\cal
O}\,. \nonumber
\ea
This implies that the identification  of the 't Hooft and Wilson operators under the action of $S$-duality
should  extend beyond   matching of their expectation values (\ref{prediccion}).
In particular, the $S$-duality conjecture relates the OPE of an 't Hooft operator to that of
the corresponding dual Wilson loop in the dual theory. Their respective OPE's are given by 
\ba
T(\LR)&=&\langle T(\LR)\rangle\left(1+
\sum_i b_i\, a^{\Delta_i} \Ocal_i
\right),
\nonumber\\
\\
W(\LR)&=&\langle W(\LR)\rangle\left(1+
\sum_i \LL c_i\, a^{\LL\Delta_i}\hskip+0.5pt\LL\Ocal_i
\right).
\nonumber
\label{OPEloopy}
\ea
Here 
$\Delta_i$ ($\LL \Delta_i$) is 
the conformal dimension of the operator $\Ocal_i$ ($\LL \Ocal_i$) and $a$ is the radius of 
the circle where the loop operators are supported.
The OPE coefficients $b_i$ and $\LL c_i$ are
non-trivial functions of the coupling constant of the theory, the choice of representation $\LR$ and the
gauge group.  

$S$-duality predicts that the 't Hooft operator   OPE  coefficient $b_i$
of  the local operator $\Ocal_i$  
of one theory
is mapped to the      Wilson operator   OPE coefficient $\LL c_i$
of the dual operator $\LL \Ocal_i$ in the dual theory.
The computation of the OPE coefficients of loop operators is closely related to the
computation of correlation functions of   loop operators and   local operators.
These  correlation functions of loop and local  operators should also transform into
each other under the action of $S$-duality.

In this paper we  compute the correlation functions of a circular 't Hooft and Wilson loop operator with an arbitrary  chiral primary operator (CPO) $\Ocal_\Delta$   in ${\cal N}=4$ super
Yang-Mills.
We show that the prediction of $S$-duality  
\ba \langle T(\LR)\cdot \Ocal_\Delta\rangle_{G,\tau}= \langle
W(\LR)\cdot \LL \Ocal_\Delta \rangle_{\LG,\Ltau}\, \label{prediccion-corr}
\ea
holds to next to leading order in the coupling constant expansion. 
This result implies that  the coefficients of chiral primary operators in the OPE of a circular 't Hooft operator at weak coupling precisely match the corresponding OPE coefficients for the dual Wilson operator   at strong coupling:
\ba
b_\Delta(\LR,\tau)= \LL c_\Delta(\LR,\Ltau).
\nonumber
\ea
Proving this   requires computing the  two point functions of chiral primary operators, which are given by free field contractions. 
We show that the two and three-point functions of chiral primary operators are invariant under the action of $S$-duality.

 In this paper we also calculate the ``scaling weight'' \cite{Kapustin:2005py} 
of the circular 't Hooft operator $T(\LR)$     at weak coupling and that of the
circular Wilson operator $W(\LR)$ at strong coupling in ${\cal N}=4$ super
Yang-Mills. This observable, which measures the conformal properties of a loop operator,   
 is determined by the OPE of the  loop operator with the stress-energy tensor. We show that 
 the scaling weight of an 't Hooft operator $T(\LR)$ evaluated at weak coupling exactly reproduces the scaling weight
 of  the dual Wilson $W(\LR)$ evaluated at strong coupling.

In summary, we   perform  novel computations with   't Hooft operators in ${\cal N}=4$ SYM  and   explicitly demonstrate  the conjectured action of $S$-duality on these observables for arbitrary gauge group $G$. This provides a quantitative demonstration of the action of electric-magnetic duality on   correlation functions in ${\cal N}=4$ SYM.

The plan of the rest of the paper is as follows.
In the next section we describe the OPE of loop operators, the notion of the scaling weight of a loop operator,
and the construction of chiral primary operators in ${\cal N}=4$ super
Yang-Mills with gauge group $G$. 
We also spell out the $S$-duality map for chiral primary operators \cite{Intriligator:1998ig,Argyres:2006qr}.
In section \ref{sec-thooft}, we compute in perturbation theory the correlation function of an 't Hooft operator with an arbitrary  chiral primary operator ${\cal O}_\Delta$ as well as the scaling weight of a circular  't Hooft operator.
Section \ref{sec-wilson} is devoted to calculating in the strong coupling expansion the correlation function of a  Wilson loop operator with ${\cal O}_\Delta$ as well as the scaling weight of a circular Wilson loop operator. These calculations are performed by solving a matrix model.
In section \ref{sec-compare} we explicitly demonstrate the    $S$-duality conjecture relating
the 't Hooft and Wilson loop correlation functions by  comparing our results
for the 't Hooft and Wilson loop correlators.
Appendix \ref{sec-conformalfactor} discusses the Weyl transformation
between $\R^4$ and $AdS_2\times S^2$, while Appendix \ref{generators} provides examples of the construction of chiral primary operators for gauge group $G$.
 Appendix \ref{app-normal} extends the equivalence of
complex  and normal matrix models for general gauge group $G$.
In Appendix \ref{2point}, we show that the two and three-point functions
of chiral primary operators are invariant under the action of $S$-duality.

\section{Loop operator OPE and  \texorpdfstring{$S$}{S}-duality}
\label{sec-OPE}
A loop operator   can be expanded in a series of local operators when   probed from a distance much larger than the characteristic size of the loop. This defines the operator product expansion (OPE) of the loop operator \cite{Shifman:1980ui,Berenstein:1998ij}.  For an operator $L$ supported on a circle of radius $a$ --   a circular  't Hooft or   Wilson   operator -- the operator product expansion is 
given by
\ba
L=\langle L\rangle\left(1+\sum_i {\cal C}_i\, a^{\Delta_i} {\cal O}_i(0)\right)\,,
\label{OPE}
\ea
where ${\cal C}_i$ is an OPE coefficient, ${\cal O}_i(0)$ is a local operator inserted at the center of the loop  and 
$\Delta_i$ is its  conformal dimension. The sum in (\ref{OPE}) is over  all conformal primary operators in the theory as well as over the   associated conformal descendant operators.

The OPE coefficients of conformal primary operators can be obtained from
the   correlation function
of the loop operator $L$ with the     primary operators  ${\cal O}_i$ 
\ba
\langle L\cdot {\cal O}_i(x)\rangle\,
\label{correla}
\ea
using the matrix of two-point functions $\langle \Ocal_i \Ocal_j\rangle$.
These correlation functions involving loop operators
are the main objects of study in this paper.

As we show in Appendix \ref{sec-conformalfactor}, super Yang-Mills in the presence of a 
  circular loop operator on $\R^4$ is Weyl equivalent to    super
Yang-Mills on $AdS_2\times S^2$ 
with the loop operator   inserted on the boundary of $AdS_2$
(the Poincar\'e disk).
By symmetry the correlator is independent of the position of the local operator
on $AdS_2\times S^2$. 
Weyl invariance of ${\cal N}=4$ super Yang-Mills  then determines the position dependence of the correlator
on $\R^4$:
\ba
\f{\langle L \cdot {\cal O}_i(x)\rangle}
{\langle L\rangle}=
\f{\Xi_i}{\tilde r^{\Delta_i}}\,.
\label{correlacircleold}
\ea
The coupling $\Xi_i$ captures the dynamical information of the correlator and our goal is to compute $\Xi_i$ for the circular
  't Hooft and Wilson loop operators in ${\cal N}=4$ super Yang-Mills with an arbitrary gauge group $G$. The conformally invariant distance  
$\tilde r$ is given by
\ba
\tilde r={\sqrt{(r^2+x^2-a^2)^2+4a^2x^2}\over2a}\,,
\nonumber
\ea
which combines  the radius $a$ of the circle where the loop operator is supported, the radial position  $r$ of the local
operator in the plane containing the loop, and the position  $x$  of the local operator in   the plane transverse
to the circle.  The OPE coefficients are most easily extracted by setting $r=0$ and expanding (\ref{correlacircleold}) in powers of $a/x$. The leading order term in this expansion of the correlator measures the OPE coefficient of the conformal primary operator ${\cal O}_i$ while the rest of the terms in the $a/x$ expansion capture the OPE coefficients of the conformal descendants of ${\cal O}_i$.

An operator that plays a central role in a conformal field theory is the stress-energy  tensor $T_{\mu\nu}$. The correlation function of a loop operator with the stress-energy tensor  measures how the loop operator transforms under a conformal transformation, and generalizes the   familiar notion of conformal dimension of a local operator to a non-local operator. This information is encoded in the ``scaling weight'' of the loop operator \cite{Kapustin:2005py}, which we   also compute for an 't Hooft and Wilson operator in ${\cal N}=4$ super Yang-Mills with gauge group $G$.

A loop operator $L$ supported on a circle of radius $a$ preserves an $SL(2,\R)\times SU(2)$ subgroup of the $Spin(1,5)$ conformal group. Conformal invariance also completely fixes the  position dependence of the correlator of the loop operator with the stress-energy tensor.  When the theory is Weyl transformed from $\R^4$ to  $AdS_2\times S^2$  in order to make the symmetries of the circular loop operator manifest (see Appendix \ref{sec-conformalfactor} for more details), the correlator of the loop operator $L$ with the stress-energy tensor     is   given
 by  \cite{Kapustin:2005py,Gomis:2008qa}
\ba
{\langle L\cdot T_{\mu\nu}(x)dx^\mu dx^\nu \rangle\over \langle L\rangle} =h_L\left(ds^2_{AdS_2}-ds^2_{S^2}\right)+{a\over 8 \pi^2}\left(ds^2_{AdS_2}+ds^2_{S^2}\right)\,,
\label{scaleweight}
\ea
 where $h_L$ is the scaling weight of the loop operator $L$,
and
the metrics on $AdS_2$ and $S^2$
are denoted by
$ds^2_{AdS_2}$ and $ds^2_{S^2}$  respectively. The last term in  (\ref{scaleweight})  captures  the conformal anomalies of the field theory   on the $AdS_2\times S^2$ geometry. For  ${\cal N}=4$ super Yang-Mills with gauge group $G$, the anomaly coefficients are      $a=c={\rm dim}(G)/4$, where ${\rm dim}(G)$ is the dimension of the gauge group.

\medskip
\medskip
\noindent
{\it Chiral primary operators and $S$-duality}
 \medskip
 
An interesting class of local operators with which to probe an 't Hooft or Wilson loop operator in
 ${\cal N}=4$ super Yang-Mills are the 
chiral primary operators.
The  $PSU(2,2|4)$ superconformal algebra implies that chiral primary operators of conformal dimension $\Delta$   belong to a  multiplet    transforming in an  $SU(4)_R$ representation with Dynkin label 
 \ba
[0,\Delta,0].
\nonumber
\ea
 In terms of the R-symmetry group
$SO(6)\simeq SU(4)_R$, these operators transform in the
 rank-$\Delta$ symmetric traceless representation of
$SO(6)$. Without loss of generality, we   
 consider the highest weight vector in the $[0,\Delta,0]$ multiplet carrying charge $\Delta$ under  the $U(1)_R$ subgroup for  which the  complex scalar field in the ${\cal N}=4$  vector multiplet
 \ba
Z\equiv \phi_1+i\phi_2 
\nonumber
\ea
is the only one charged.\footnote{The other operators in the $[0,\Delta,0]$ 
multiplet for any $\Delta$ take the form $C^{i_1\ldots i_\Delta} K_{a_1\ldots a_\Delta}
\phi^{a_1}_{i_1}\ldots \phi^{a_\Delta}_{i_\Delta}$,
where $C^{i_1\ldots i_\Delta}$ is a symmetric traceless tensor
and $K_{a_1\ldots a_\Delta}$ is defined by
$P(Z)=K_{a_1\ldots a_\Delta} Z^{a_1}\ldots Z^{a_\Delta}$.
} 
 
Chiral primary operators involving only $Z$
are  given by  $G$-invariant polynomials  of $Z$, and  form a ring. The ring multiplication law is the usual operator product. These operators are part of the usual ${\cal N}=1$ chiral ring, and are the lowest components of chiral superfields with respect to a particular ${\cal N}=1$ subalgebra of ${\cal N}=4$.    This ring has 
as many generators as the rank $r$
of the   group $G$.  Let us denote the  polynomials  
 generating the $G$-invariant ring by
 \ba
P_{1}(Z), P_{2}(Z),\ldots, P_{r}(Z)\,.
\label{ring-gen}
\ea
The degrees  $\{\nu_i\}=\{\nu_1,\ldots, \nu_r\}$ of these polynomials $P_i(Z)$
are the set of positive integers 
that appear as the order of the Casimirs of $G$.\footnote{More
precisely $\nu_i$ are the order of those Casimirs
which generate the center of the universal enveloping algebra.
The integers $\nu_i-1$ are known as the {\it exponents} of $G$.
}

In a  group $G$ admitting a Casimir of order $\nu$, there exists    a rank-$\nu$   {\it invariant} symmetric tensor on the Lie algebra, which we denote by $K_{a_1\ldots a_\nu}$, where $a_i=1,\ldots, {\rm dim}(G)$. 
Each   generator of the chiral ring (\ref{ring-gen}) can  be written in terms of such a  tensor as
 \ba
 P(Z)=K(Z,\ldots,Z)=K_{a_1\ldots a_\nu}Z^{a_1}\ldots Z^{a_\nu}\,,
 \label{CPOgeneralnew}
 \nonumber
 \ea
where  $Z\equiv Z^a T_a$, and $T_a$ are the generators of the Lie algebra. We list the generators of the ring for several choices 
of $G$ in Appendix \ref{generators}.

The most general chiral primary operator constructed from $Z$ is then given by
\ba
\Ocal_\Delta\equiv \oo{g^\Delta}P_\Delta(Z)\,,
\label{CPOgeneral}
\ea
where\footnote{In order to not clutter notation we do not make explicit the dependence of the operator on $\{N_i\}$.} 
\ba
P_\Delta(Z)\equiv\prod_{i=1}^r P_{i}(Z)^{N_i}\,,
\label{P-prod}
\ea
and $N_i$ are non-negative integers. 
In our convention 
 a chiral primary
operator   (\ref{CPOgeneral}) has an explicit coupling constant dependence, while
 the polynomials $P_{i}(Z)$ do not depend on the coupling
as explained in Appendix \ref{generators}.
The conformal dimension of the chiral primary operator (\ref{CPOgeneral}) is given by
\ba
\Delta=\sum_{i=1}^r N_i \nu_i\,.
\nonumber
\ea
Therefore the spectrum of conformal dimensions
is determined by the order of the Casimirs of $G$:
\begin{table}[htbp]
\begin{center}
\begin{tabular}{c||c}
Group $G$ &Order $\nu$ of Casimirs
\\
\hline
\hline
$A_{n-1}=SU(n)$ & $2,3,\ldots, n$\\
\hline
$B_n=SO(2n+1)$ & $2,4,\ldots, 2n$
\\
\hline
$C_n=Sp(n)$ & $2,4,\ldots, 2n$
\\
\hline
$D_n=SO(2n)$ & $2,4,\ldots, 2n-2, n$
\\
\hline
$E_6$ & $2,5,6,8,9,12$
\\
\hline
$E_7$ & $2,6,8,10,12,14,18$
\\
\hline
$E_8$ & $2,8,12,14,18,20,24,30$
\\
\hline
$F_4$ & $2,6,8,12$
\\
\hline
$G_2$ & $2,6$
\end{tabular}
\end{center}
\begin{center}
{\rm Table 1. Casimirs for simple Lie groups.}
\end{center}
\end{table}

We are interested in the behaviour of chiral primary operators under the action of $S$-duality,   which  exchanges the gauge group $G$ with the  dual gauge group $\LG$
\ba
G\longleftrightarrow \LG\,.
\nonumber
\ea
As in \cite{Argyres:2006qr} we use
the metric,
normalized so that short coroots have length $\sqrt 2$,
to identify the Cartan subalgebra
of each group with its dual vector space,
denoted by $\tfrak$ and $\Lt$ for $G$ and $\LG$
respectively.\footnote{Here we use the metrics on $\tfrak$ and $\Lt$
to identify $\tfrak$ with $\tfrak^*$ and $\Lt$ with $\Lt^*$,
respectively, while we make explicit
the isomorphism $\Rcal:\tfrak\ra \Lt$.
In \cite{Donagi:2006cr} and \cite{Gukov:2006jk} another
convention was used where $\tfrak^*$ and $\Lt$ were
taken to be equal, while the isomorphisms 
$\tfrak \ra \tfrak^*$ and $\Lt \ra \Lt^*$,  constructed 
using the metrics 
were made explicit.}
For the dual groups $G$ and $\LG$, 
there is by definition a linear transformation
\ba
\tfrak \ra \Lt
\label{lineartransform}
\ea
that maps roots of $G$ to coroots of $\LG$.
We denote this map by $n_\gfrak^{1/2} \Rcal$, where 
   $n_\gfrak=1,2$ or $3$ is the ratio of the length-squared
of the long and short roots in the Lie algebra $\gfrak$ and 
$\Rcal$ is a norm-preserving linear transformation. The transformation
  $n_\gfrak^{1/2} \Rcal^\mo$ maps roots of $\LG$ to coroots of $G$. The
  transformation $\Rcal$ is unique up to the action of the Weyl group \cite{Goddard:1976qe}.
  For simply laced groups, $\Rcal$ can be taken to be the identity operator.

The conjecture \cite{Intriligator:1998ig,
Intriligator:1999ff, Argyres:2006qr}  is that the ring generators,
and therefore all operators  in the chiral ring  for gauge groups  $G$ and $\LG$ are mapped into each other under the action of $S$-duality.
The precise proposed mapping is
\ba
\Ocal_\Delta=\oo{g^\Delta}P_{\Delta}(Z)\longleftrightarrow
\LL \Ocal_\Delta=\oo{\Lgc^\Delta} \LL P_{\Delta}(\LL Z)\,,
\label{Sact}
\ea
where the $\LG$-invariant polynomial   $\LL P_\Delta$  of 
$\LL Z\in \LL\gfrak_\C$ is uniquely determined by $P_\Delta$ through
the relation 
\ba
P_\Delta(\lambda)= \LL P_\Delta({\cal R} \lambda),~~~\forall \lambda \in \tfrak.
\label{P-LP}
\ea
The conjectured action of $S$-duality on chiral primary operators (\ref{Sact})  is consistent with the mathematical fact that $\{\nu_i(G)\}=\{\nu_i{(\LG})\}$. For all gauge groups, $G$ and $\LG$ share the same Lie algebra except for $SO(2n+1)$ and $Sp(n)$.  Their Lie algebras are exchanged under $S$-duality and have the same set of orders for
Casimirs as seen in Table 1.
Other
For $G=U(n)=\LG$, the map (\ref{Sact}) is simply  given by $g^{-\nu}\tr Z^\nu
\leftrightarrow (\Lgc)^{-\nu} \tr (\LL Z)^{\nu}$.
See Appendix \ref{generators} for more details on the $S$-duality
map of chiral primary operators.

We note that for any choice of gauge group $G$ there is a universal  $\Delta=2$ chiral primary operator
\ba
\Ocal_2=\oo{g^2}\tr\, Z^2\,,
\nonumber
\ea
where $\tr(\cdot~\cdot)$
 is the invariant quadratic form on $\gfrak$
whose restriction to $\tfrak$ is the metric on the subalgebra.
It was shown in \cite{Gomis:2008qa}
using supersymmetric Ward identities that the correlator of a circular loop operator $L$ with $\Ocal_2$  can be related to the correlator (\ref{scaleweight}) of the same circular loop operator with the stress-energy  tensor $T_{\mu\nu}$, which also universally exists for any choice of $G$. This allows us to compute the scaling weight $h_L$ of a  circular  't Hooft and Wilson  loop operator   in ${\cal N}=4$ super Yang-Mills with gauge group $G$ in terms of the conformal dimension two chiral primary operator coupling $\Xi_{2}$  (\ref{correlacircleold})   using the formula 
\cite{Gomis:2008qa}
\ba
h_L=-\f{4}{3} \Xi_{2}\,.
\label{wardiden}
\ea
%

\section{Quantum 't Hooft loop correlators}
\label{sec-thooft}

In this section we compute the correlation function of a circular 't Hooft  operator 
with an arbitrary chiral primary operator ${\cal O}_\Delta(Z)$   in ${\cal N}=4$ super Yang-Mills with gauge group $G$. We give explicit formulas for $\Xi_{\Delta}$   (\ref{correlacircleold}) and for the scaling weight $h_T$ 
of the circular 't Hooft operator  to next to leading order in the weak coupling expansion. 
Before delving into the details of these computations we first give a minimal discussion of 't Hoof operators in ${\cal N}=4$ super Yang-Mills.

An 't Hooft operator  inserts a magnetically charged source into the theory. In a theory with gauge group $G$ 
an 't Hooft operator is labeled \cite{Kapustin:2005py}  by a representation $\LR$ of the dual group  $\LG$ \cite{Goddard:1976qe}. We denote an 't Hooft operator  labeled by a representation $\LR$ by $T(\LR)$.

A circular loop operator in a conformal field theory in $\R^4$ preserves an   $SU(1,1)\times SU(2)$ group of symmetries. Explicit computations with a circular 't Hooft operator  $T(\LR)$ are  conveniently performed by conformally mapping  the theory from $\R^4$ to $AdS_2\times S^2$, where   $AdS_2$ 
is modeled by the Poincar\'e disk. 
 The   symmetries preserved by the circular 't Hooft operator  are made manifest  in  $AdS_2\times S^2$, as they act by  isometries.
In  $AdS_2\times S^2$, the loop operator is supported at the conformal  boundary of $AdS_2\times S^2$, identified with the circular boundary of the  Poincar\'e disk.

The insertion of an 't Hooft operator $T(\LR)$ at the conformal boundary of $AdS_2\times S^2$ creates the following classical field configuration \cite{Kapustin:2005py}
\ba
F^0= \f{B}2 {\rm vol}(S^2)+i g^2\theta {B\over 16\pi^2}
{\rm vol}(AdS_2),~~~\phi^0_1=\f{B}{2}{g^2\over 4\pi} |\tau|\,.
\label{theta-background}
\ea
The coefficient $B\equiv B^i {H_i} \in
\tfrak$ takes values in the Cartan subalgebra of the Lie algebra
$\gfrak$ associated with the gauge group $G$.  
Via (\ref{lineartransform}) $B$ can be identified \cite{Goddard:1976qe} 
with the highest weight $\Lw$
of a  representation $\LR$ of the dual group $\LG$, justifying the
labeling of 't Hooft operators in terms of representations of the
dual group \cite{Kapustin:2005py}. The insertion of an 't Hooft operator creates  quantized 
magnetic field,  and when
$\theta\neq 0$ it also generates an electric field, as the monopole that is being inserted 
acquires electric charge via the Witten effect \cite{Witten:1979ey}.
Without loss of generality, we have chosen the single scalar field  that is excited by the circular 't Hooft operator to be  $\phi_1$.

In order to compute the correlation function of  $T(\LR)$ with a chiral primary operator ${\cal O}_\Delta(Z)$   a quantum definition of the 't Hooft operator is required. This quantum definition was proposed  in \cite{Gomis:2009ir}, where it was used to explicitly compute the expectation value of $T(\LR)$ to next to leading order in perturbation theory and to exhibit the conjectured action of $S$-duality on circular 't Hooft and Wilson operators in 
${\cal N}=4$ super Yang-Mills.\footnote{The paper \cite{Kapustin:2007wm} considered
semiclassical quantization of 't Hooft line operators
in a holomorphic-topological twisted
version of $\Ncal=4$ super Yang-Mills and obtained
the associated Hilbert spaces by calculating the zero-modes around
the background field configuration.
}

The basic proposal in \cite{Gomis:2009ir} is to define the gauge invariant 't Hooft operator by a path integral quantized in the background field gauge expanded around the background  (\ref{theta-background})
\ba 
A&=A^0+\hat{A}\,,
\cr \phi_I&=\phi^0_I+\hat{\phi}_I\,. \nonumber 
\ea
 In this path integral   one must integrate over all quantum fields (gauge fields, scalars, fermions and ghosts) with the boundary conditions specified by  (\ref{theta-background}).\footnote{The definition of
the 't Hooft operator in terms of  an   ${\cal N}=4$ super Yang-Mills  
  partition function  on $AdS_2\times S^2$
is reminiscent of Sen's   definition of the quantum entropy
function \cite{Sen:2008yk, Gupta:2008ki, Sen:2008vm, Sen:2009vz, Banerjee:2009af} in terms of  the 
string theory  path  integral  on $AdS_2$, which  encodes  the macroscopic degeneracy of states of extremal
black holes. It would be interesting to understand whether a direct physical relation between the two path integrals exists.}  The classical field configuration (\ref{theta-background}) created by the 't Hooft operator $T(\LR)$ breaks the $G$-invariance of the theory to invariance under an stability group $H\subset G$.
 The
choice of $B\in \tfrak$, which characterizes the background, determines the unbroken gauge group $H$. This is
generated by those $x \in \gfrak$ for which
\ba
[x,B]=0\,.
\label{levy}
\ea
In order to have a path integral definition of the 't Hooft operator
$T(\LR)$ which is gauge invariant, we must integrate over the
$G$-orbit of $B\in \tfrak$ along the loop. This integration, which
we include in our definition of the path integral measure, restores
$G$-invariance. The integral we must perform is over the adjoint
orbit of $B$
\ba
 O(B)= \{\mathsf{g} B\mathsf{g}^\mo, \ \mathsf{g} \in G\}\,,
 \label{adjoint}
\ea
 We refer the reader to  \cite{Gomis:2009ir}  for more details on the path integral definition of an 't Hooft operator.

 Using the path integral prescription in \cite{Gomis:2009ir}, we now proceed to compute the correlator of an 't Hooft operator $T(\LR)$ with an arbitrary chiral primary operator ${\cal O}_\Delta(Z)$ in ${\cal N}=4$ super Yang-Mills with gauge group $G$.  When the theory is defined on $AdS_2\times S^2$, conformal invariance implies that the correlator is given by 
\ba
\f{\langle T(\LR) \cdot \Ocal_\Delta\rangle_{G,\tau}}
{\langle T(\LR)\rangle_{G,\tau}}={\Xi_{\Delta}}\,,
\label{correlaAdS}
\ea
where $\Xi_{\Delta}$ is a function that depends on the representation $\LR$ of the 't Hooft operator, the  complexified coupling constant $\tau$ and the choice of gauge group $G$.

We evaluate this correlator  by expanding    the path integral  representation of the correlator (\ref{correlaAdS}) around the classical field configuration (\ref{theta-background}) created by the 't Hooft operator $T(\LR)$. To next to leading  order  in perturbation theory it suffices to expand the gauge fixed ${\cal N}=4$ super Yang-Mills action  and the   operator insertion ${\cal O}_\Delta(Z)$ to quadratic order in the fluctuations. We then proceed to integrate over
the quantum fluctuations at one loop.

  The chiral primary operator $\Ocal_\Delta=g^{-\Delta}P_\Delta(Z)$ can be  expanded around the background (\ref{theta-background}) 
  by decomposing the complex scalar field $Z$ in a basis of  Lie algebra generators through $Z=Z^aT_a$, where $a=1,\ldots {\rm dim}(G)$. To quadratic order in the fluctuations we have\footnote{This correlator with an 't Hooft operator replaced by a surface operator \cite{Gukov:2006jk} (see also \cite{Gomis:2007fi}) was evaluated in the  leading semiclassical approximation in \cite{Drukker:2008wr}.}
\ba
  \Ocal_\Delta=\hskip-3pt\left({g|\tau|\over 8\pi}\right)^\Delta\hskip-3pt\left[P_\Delta(B)+ \f{8\pi}{g^2|\tau|}\hat Z^a  \p_a P_\Delta(B)+
  \half\left(\f{8\pi}{g^2|\tau|}\right)^2 \hat Z^a\hat Z^b\p_a \p_b P_\Delta(B)\right],
\ea
  where we have used that  $P_\Delta(Z)$ given in (\ref{P-prod})
 is a polynomial of degree $\Delta$. We note that the scalar field $Z=\phi_1+i\phi_2$ involves a scalar field $\phi_1$ that is excited in the 't Hooft operator background (\ref{theta-background}) and another one $\phi_2$ that is not.

The correlator to next to leading order in perturbation theory is then given by
\ba
{\langle T(\LR) \Ocal_\Delta\rangle_{G,\tau}\over \langle T(\LR) \rangle_{G,\tau}}=
\left({g|\tau|\over 8\pi}\right)^\Delta\left[ P_\Delta(B)+\half\left(\f{8\pi}{g^2|\tau|}\right)^2 \p_a \p_b P_\Delta(B)\langle
{\hat \phi}_1^a \hat\phi_1^b-\hat \phi_2^a \hat\phi_2^b\rangle\right]\,,
\label{thooft-parametrization}
\ea
where $\langle
\hat \phi_1^a \hat\phi_1^b-\hat \phi_2^a \hat\phi_2^b\rangle$ is the difference between the scalar propagator for $\hat \phi_1$ and   $\hat \phi_2$ in the 't Hooft operator background (\ref{theta-background}). In arriving at (\ref{thooft-parametrization}) we have used that $\langle \hat Z^a\rangle=0$ as well as  $\langle \hat\phi_1 \hat\phi_2\rangle$=0, 
which follows from $SO(5)$ invariance of the 't Hooft operator background (\ref{theta-background}).

 The first term in (\ref{thooft-parametrization}) is the leading semiclassical approximation,
where the chiral primary operator is evaluated on the classical field configuration (\ref{theta-background}). The second term is the one loop correction. At one loop we must sum over all possible contractions between two fields in the operator $P_\Delta(Z)$, while the remaining  $\Delta-2$  scalar fields in the operator are to be evaluated on the classical background (\ref{theta-background}). The second term in (\ref{thooft-parametrization}) sums over all possible contractions between two scalar fields, which are connected by the scalar field propagator on the 't Hooft operator background.
What we need is the difference of propagators
\ba
\langle \hat \phi_1^a \hat \phi_1^b
-
 \hat \phi_2^a \hat \phi_2^b
\rangle
=
\langle \hat \phi_1^a \hat \phi_1^b
-
 \hat \phi_2^a \hat \phi_2^b
\rangle_0
+
\langle \hat \phi_1^a \hat \phi_1^b
-
 \hat \phi_2^a \hat \phi_2^b
\rangle_{0\hskip-4.4pt /}. \label{zero-nonzero}
\ea
where all the fields are evaluated at the same spacetime point.
On the right hand side we have separated the contributions
of zero modes from those of non-zero modes.

We now argue that the second term in the right hand side of
(\ref{zero-nonzero}) vanishes, {\it i.e.},
the non-zero modes cancel out in the difference of propagators.
When we introduce an IR cut-off to discretize the spectrum,
the  second term 
takes the form
\ba
\langle \hat \phi_1^a \hat \phi_1^b
-
 \hat \phi_2^a \hat \phi_2^b
\rangle_{0\hskip-4.4pt /}
=\sum_n \oo{\lambda_n} f^a_{1n}(x)  f^b_{1n}(x)
-\sum_m \oo{\omega_m} f^a_{2m}(x)  f^b_{2m}(x).
\label{eigen}
\ea
Here $f_{2m}^a(x)$ is the normalized eigenfunction
of the scalar Laplacian in the background (\ref{theta-background}) with eigenvalue $\omega_m$. This is the 
linearized operator for fluctuations of  the scalar field $\hat \phi_2^a$.
For $\hat\phi_1^a$,  note that the quadratic terms
in the gauge-fixed action mix $\hat\phi_1^a$ with gauge the field fluctuations (see \cite{Gomis:2009ir} for the precise form of the gauge fixed action).
Thus  $f_{1n}^a$ is a component of the vector-valued
eigenfunction for the relevant differential operator
with eigenvalue $\lambda_n$.
The eigenfunctions $f_{1n}^a$ and $f_{2m}^a$
are non-constant, since they are non-zero modes.
On the other hand, the  symmetries of $AdS_2\times S^2$
dictate that the total expression (\ref{eigen}), which is finite, has to be constant
in the limit that the IR cut-off is removed.
This implies that the non-zero modes of $\hat\phi_1^a$ and $\hat \phi_2^a$
have to cancel out in (\ref{eigen}) in the limit that the regulator is removed, and therefore
we can drop the second term in the right hand side of (\ref{zero-nonzero}) and focus on the zero-mode contribution.

We  now proceed to show that zero modes, which are constant,
 give a non-trivial  contribution to the 
 correlation functions.
As we have already mentioned, the background (\ref{theta-background}) created by the insertion of an 't Hooft operator $T(\LR)$ breaks the gauge group $G$ down to a subgroup $H$. It was argued in  \cite{Gomis:2009ir} that  in order to make the 't Hooft operator $T(\LR)$ gauge invariant   one must integrate over the $G$-adjoint orbit of $B$ (\ref{adjoint}), obtained by 
the action of $G$ on the classical background  (\ref{theta-background}).
Conjugating the scalar classical background  (\ref{theta-background}) generates quantum fluctuations which are  associated with zero modes of the quadratic operator for $\hat \phi_1$. The fluctuations generated by a $G$-transformation are given by
\ba
\hat\phi_1=
\delta B \f{g^2}{8\pi}|\tau|\equiv
 i[\xi, B]\,{g^2\over 8\pi} |\tau|~~~~~~~~~~~~~ \xi\in\gfrak\,.
\label{zeromode0}
\ea
We can identify the non-vanishing fluctuations by writing the Lie algebra $\gfrak$ in the Cartan basis $\{{H_i}, {E_{\alpha}}\}$, where the generators ${H_i}$ span the Cartan subalgebra $\tfrak\subset \gfrak$ and ${E_{\alpha}}$ are ladder operators associated to  roots $\alpha$ of the Lie algebra $\gfrak$.
In this basis $\xi$ takes the form $\xi=\xi^iH_i+\xi^\alpha E_\alpha$. Since $B=B^i {H_i}$ is in the Cartan subalgebra we have that the non-vanishing scalar field fluctuations are 
\ba
\hat\phi_1= \sum_{\alpha(B)\neq 0}  
\alpha(B) \xi^\alpha
E_\alpha
{g^2\over 8\pi} |\tau|\,,
\label{zeromode}
\ea
where we have used the commutation relation $[\lambda,E_\alpha]=\alpha(\lambda) E_\alpha$, valid  for any $\lambda\in \tfrak$.
The sum in  (\ref{zeromode}) is over all the roots $\alpha$ 
that do not annihilate $B$, as those which do annihilate $B$ 
do not contribute. This implies that these fluctuations 
(\ref{zeromode}) are labeled by  the coset space $G/H$, where $H\subset G$ is the subgroup that preserves the field configuration  (\ref{theta-background}) created by the  't Hooft operator $T(\LR)$.
This follows from the definition of $H$ given in (\ref{levy}), which is generated in the Cartan basis by $\{E_{\alpha}| \alpha(B)\neq 0\}$.
Therefore, the coset space $G/H$ parametrizes the space of zero mode fluctuations of the scalar field $\hat\phi_1$.

The path integral representation of the correlation function (\ref{correlaAdS}) is gauge invariant
 once we integrate over the zero mode fluctuations of the scalar field $\hat\phi_1$ obtained from the classical background (\ref{theta-background}) 
  by  the action of $G$. The integration measure for these modes follows from the quadratic form defined by the ${\cal N}=4$ super Yang-Mills on-shell action evaluated on the 't Hooft operator background (\ref{theta-background}). We recall that the renormalized, on-shell action is given by  \cite{Gomis:2009ir}
\ba
  S= {\tr(B^2) \over 8}
g^2|\tau|^2 \,,
\label{circleresult}
\ea
and defines the quadratic form from which the propagator can be computed. We first note that  fluctuations of $B$ along root directions can be expanded as $\delta B=\sum_{\alpha>0}(\delta B^\alpha E_\alpha+\delta B^{-\alpha} E_{-\alpha})$.
Using the on-shell action (\ref{circleresult})  we get that 
\ba
\langle \delta B^\alpha  \delta B^{-\alpha}\rangle={2|\alpha|^2\over g^2|\tau|^2}
\label{propsB}
\ea
where we have used that $\tr E_\alpha E_{-\alpha}=2/|\alpha|^2$, and where $|\alpha|^2=\langle \alpha,\alpha\rangle$ is the length of the root  $\alpha$ computed using the restriction of the  metric on $\gfrak$ to the Cartan subalgebra $\tfrak$.
The propagator for the scalar field fluctuations is given by
\ba
 \langle    \hat \phi^\alpha_1~\hat \phi^{-\alpha}_1 \rangle_0 =\langle \delta B^\alpha  \delta B^{-\alpha}  \rangle \left(g^2 |\tau|\over 8\pi\right)^2\,.
 \nonumber
 \ea
Therefore, using (\ref{propsB}) we arrive at
\ba
 \langle  \hat \phi^\alpha_1~  \hat \phi^{-\alpha}_1 \rangle_0 = {g^2\over 32\pi^2}|\alpha|^2\,.
 \label{propagascalar}
 \ea

Since only the zero-modes of $\hat\phi_1$ contribute,
(\ref{thooft-parametrization}) simplifies to
\ba
{\langle T(\LR)\cdot \Ocal_\Delta(Z)\rangle_{G,\tau}\over \langle T(\LR) \rangle_{G,\tau}}=
\hskip-3pt\left({g|\tau|\over 8\pi}\right)^\Delta\hskip-3pt\left[ P_\Delta(B)+\left(\f{8\pi}{g^2|\tau|}\right)^2\hskip-10pt\mathop{\sum_{\alpha>0}}_{\alpha(B)\neq 0}
\p_\alpha\p_{-\alpha} P_\Delta(B)\langle
 \hat \phi_1^{\alpha}  \hat\phi_1^{-\alpha}\rangle_0\right],
\nonumber
\ea
and using (\ref{propagascalar}) we obtain
\ba
{\langle T(\LR) \cdot \Ocal_\Delta\rangle_{G,\tau}\over \langle T(\LR) \rangle_{G,\tau}}=
\left({g|\tau|\over 8\pi}\right)^\Delta\left[ P_\Delta(B)+
\f{2}{g^2|\tau|^2}\mathop{\sum_{\alpha>0}}_{\alpha(B)\neq 0}{|\alpha|^2}\p_\alpha\p_{-\alpha} P_\Delta(B)\right]\,.
\label{thooft-parametrizationa2}
\ea
By using the relation\footnote{This can be
shown by expanding the equation $P(\sfg Z \sfg^\mo)=P(Z)$
with $\sfg=\exp\left(i \xi^i H_i+ i\xi^\alpha E_\alpha\right)$ for
small $\xi$.}
 \ba
\hat\alpha\cdot \p P_\Delta(\lambda)
\equiv \hat\alpha^i \p_i P_\Delta(\lambda)
=\alpha(\lambda)\p_\alpha\p_{-\alpha} P_\Delta(\lambda),
~~~~\forall \lambda \in \tfrak_\C,
 \label{inv-rel1}
 \ea
where $\hat\alpha=[E_\alpha, E_{-\alpha}]=2\alpha/|\alpha|^2$
is the coroot corresponding to $\alpha$,
we can further rewrite the
correlator as
\ba
{\langle T(\LR) \cdot \Ocal_\Delta\rangle_{G,\tau}\over \langle T(\LR) \rangle_{G,\tau}}=
\left({g|\tau|\over 8\pi}\right)^\Delta\left[ P_\Delta(B)+\f{2}{g^2|\tau|^2}
\mathop{\sum_{\alpha>0}}_{\alpha(B)\neq 0}
\f{\langle \alpha,\alpha\rangle}{\alpha(B)} \hat\alpha\cdot \p 
 P_\Delta(B)\right]\,.
\label{thooft-parametrizationa3}
\ea
This is the final result  to next to leading order in perturbation theory for the  correlator of an 't Hooft operator $T(\LR)$ and an arbitrary chiral primary operator $\Ocal_\Delta$    in  ${\cal N}=4$ super Yang-Mills with   gauge group $G$.
  
As illustration of the general result (\ref{thooft-parametrizationa3}),
let us consider the case with gauge group $G=U(n)$, for chiral primary operator $P_\Delta=\tr Z^\Delta$
and for 't Hooft operator labeled by the highest weight $B={\rm diag}(m_i)$ 
with $m_1>m_2\ldots>m_n$.  In this case, the
correlation function is given by
\ba
\f{\langle T([m_1,m_2,\ldots,m_n])\cdot \Ocal_\Delta\rangle_{G,\tau}}
{\langle T([m_1,m_2,\ldots,m_n])\rangle_{G,\tau}}
=\left(\f{g|\tau|}{8\pi}\right)^\Delta
\left[
\sum_i m_i^\Delta+\f{4\Delta}{g^2|\tau|^2}\sum_{i<j}
\f{m_i^{\Delta-1}-m_j^{\Delta-1}}{m_i-m_j}
\right].
\nn
\ea

Using the formula (\ref{wardiden}),
that follows from a supersymmetric Ward identity \cite{Gomis:2008qa}, 
we can obtain the scaling weight of an arbitrary 't Hooft operator $T(\LR)$ 
from
the correlator of the 't Hooft operator  with the $\Delta=2$ chiral primary operator. The one loop expression for the scaling weight  of an  't Hooft operator  in ${\cal N}=4$ super Yang-Mills for an arbitrary gauge group $G$ is given by
 \ba
 h_T(\LR,\tau)=-
{g^2|\tau|^2\over 48\pi^2}\left[\tr (B^2)+
\f{8}{g^2|\tau|^2}
\dim(G/H) \right]\,.
\label{scaling-result}
 \ea
The second term is the first quantum correction
to the classical computation 
considered in \cite{Kapustin:2005py}.\footnote{In the 
formula for the scaling weight of the BPS 't Hooft operator in 
\cite{Kapustin:2005py}, the sign for the gauge field
contribution should be changed.
With this modification taken into account, our leading result in (\ref{scaling-result})
is consistent with   \cite{Kapustin:2005py}.
}

\section{Wilson loop correlators at strong coupling}
\label{sec-wilson}

In this section we perform the strong coupling expansion
of the correlator of the circular Wilson loop operator \cite{Rey:1998ik,Maldacena:1998im}
\ba W(R)\equiv \Tr_R {\rm P} \exp{{\oint(iA+\phi_1)}}\,, \nonumber
\ea
and
an arbitrary chiral primary operator $\Ocal_\Delta$:
\ba
\langle W(R)\cdot  \Ocal_\Delta\rangle_{G,\tau}.
\label{correlaope}
\ea

It was first noticed in \cite{Erickson:2000af} that, in Feynman gauge, the combined propagator for the gauge field and the scalar  between two points on the circle is position-independent (also independent of the radius 
$a$ of the circle), and that  Feynman diagrams with internal vertices cancel  to leading  order in perturbation theory.  This led to the remarkable conjecture that the expectation value of a circular Wilson loop operator in ${\cal N}=4$ super Yang-Mills
is captured by a matrix integral \cite{Erickson:2000af,Drukker:2000rr}, which has now been proven in \cite{Pestun:2007rz} using localization.

In \cite{Semenoff:2001xp}, it was shown to leading order in perturbation theory that Feynman diagrams with 
internal vertices contributing to the correlator
(\ref{correlaope})
 vanish, also leading to the conjecture\footnote{The 
large $N$ conjecture for the correlators of  half BPS Wilson and local operators
has been tested extensively using AdS/CFT
 \cite{Semenoff:2001xp,Okuyama:2006jc,Giombi:2006de,Gomis:2008qa}.
Given that the finite $N$ version of the conjecture for the expectation value 
has been proven, it seems likely that the conjecture for the
correlator also holds for finite rank, and that it can   be proven using localization.  Progress in this direction has been made recently
in \cite{Pestun:2009nn,Giombi:2009ds,Bassetto:2009rt}.
 } 
that   loop corrections arising from internal vertices   cancel to all orders in perturbation theory.%
\footnote{See \cite{Semenoff:2006am} for an extension to 
the correlators of $1/4$ BPS Wilson loops and half BPS local operators.
} 
This conjecture implies that all quantum corrections to the correlator (\ref{correlaope}) are due to ladder diagrams, reducing the sum   over all Feynman diagrams  to a combinatorial problem   \cite{Semenoff:2001xp}.
This combinatorics is exactly captured by a complex 
Gaussian matrix model defined by a partition function 
where the complex matrix $z$ is an element of the complexified
Lie algebra $\gfrak_\C$ \cite{Okuyama:2006jc}. 
The same matrix model also computes \cite{Okuyama:2006jc} the two, and three-point 
functions of local chiral primary operators in 
$\N = 4$ super Yang-Mills of the form $\Ocal_\Delta=g^{-\Delta}P_\Delta(Z)$ 
as in (\ref{CPOgeneral}),
where $P_\Delta$ is a $G$-invariant polynomial (\ref{P-prod}).
Therefore, the correlator of the circular Wilson loop  
$W(R)$ with the chiral primary operator $\Ocal_\Delta$
is conjecturally given by\footnote{This is the form of the correlator when the theory is defined on $AdS_2\times S^2$. In $\R^4$ we should further divide by $\tilde{r}^\Delta$ as in (\ref{correlacircleold}).}
\ba
\langle 
W(R)\cdot
\Ocal_\Delta 
\rangle_{G,\tau}=
 \oo{(2\pi g)^\Delta}
\f{\displaystyle
 \int_{\gfrak_\C} [dz]e^{-\f{2}{g^2}\tr( \bar z z)
}
\Tr_R \, e^{\f{z+\bar z}2}
P_\Delta(z)
}{
\displaystyle
\int_{\gfrak_\C} [dz]e^{-\f{2}{g^2}\tr (\bar z z)
}
},
\ea
where $[dz]$ is the measure on the complexified Lie algebra $\gfrak_\C$.
We   use this matrix model representation   to compute the correlator (\ref{correlaope}) in the strong coupling expansion.

It is possible to rewrite this complex matrix model
as a normal matrix model, where the integration
is performed over the elements $z$ in the complexified
Lie algebra  that commute
with its  conjugate variable: $[z,\bar z]=0$.
The normal matrix integral can be further restricted to
the complexified Cartan subalgebra ${\mathfrak t}_\C$.
This is shown in Appendix \ref{app-normal}
for an arbitrary gauge group $G$, thereby
  generalizing the derivation in Appendix A of \cite{Okuyama:2006jc}, 
where the case $G=U(n)$ was studied.
The precise relation
between the complex matrix model
and the normal matrix model is given by  
\ba
&&
\f{
\displaystyle
\int_{\gfrak_\C}[dz] e^{-\f{2}{g^2} \tr(\bar z z)}
\Tr_R\, e^{\f{z+\bar z}2} P_\Delta(z)
}{
\displaystyle
\int_{\gfrak_\C}[dz] e^{-\f{2}{g^2} \tr( \bar z z)}
\Tr_R\, e^{\f{z+\bar z}2}
}
\nn\\
&&\hspace{10mm}
=
 \left(\f{g^2}4\right)^\Delta
\f{\displaystyle
\sum_{v} n(v)
e^{\f{g^2}8 \langle {v},{v}\rangle}
\int_{{\mathfrak t}_\C} [dz]
\left|\Delta\left( {v}+\f{2}g z\right)\right|^2
e^{-\langle \bar z,z\rangle}
P_\Delta\left({v}+\f{2}{g} z\right)
}{\displaystyle
\sum_{v}n(v)
 e^{\f{g^2}8 \langle {v},{v}\rangle}
\int_{{\mathfrak t}_\C} [dz]
\left|\Delta\left( {v}+\f{2}g z\right)\right|^2
e^{-\langle \bar z,z\rangle}
}\,, \label{complex-normal2}
\ea
where 
\ba
\Delta(z)
=\prod_{\alpha>0} \alpha(z)
\nonumber
\ea
generalizes the Vandermonde determinant that appears in the $G=U(n)$ case,
and  $\langle\  ,\ \rangle$ is the restriction of the metric
 $\tr(\cdot\,\,\cdot)$
to the Cartan subalgebra
${\mathfrak t}$. In order to derive (\ref{complex-normal2}) we have expressed the insertion of the
character $\Tr_R\, e^{\f{z+\bar z}2}$ 
 in the representation $R$ 
as a sum over the weights $v$ 
 in the representation $R$ of the gauge group $G$, and $n(v)$ is the multiplicity of the
weight $v$ in the representation $R$.

In the strong coupling limit, the terms
with weights $v$ in the Weyl-orbit of the highest weight $w$ in the representation $R$ dominate, as $\langle {v},{v}\rangle$ is maximal for these.
The leading term at strong couping
is simply given by $P(w)$.
To study corrections, it is convenient to split $\Delta(z)$ as
\ba
\Delta(z)=\Delta_{G/H}(z)\Delta_H(z),
\nonumber
\ea
where
\ba
\Delta_{G/H}(z)\equiv
\mathop{\prod_{\alpha>0}}_{\langle \alpha,w\rangle\neq 0}
\alpha(z),~~~
\Delta_{H}(z) \equiv
\mathop{\prod_{\beta>0}}_{\langle \beta,w\rangle= 0}
\beta(z).
\nonumber
\ea
The correction to next to leading order in the strong coupling expansion, where $g\gg 1$, 
comes from the contraction of $z^i\p_i P_\Delta\equiv z\cdot \p P_\Delta$
with $\bar z\cdot \p\Delta_{G/H}$.\footnote{The
contraction of $z\cdot \p P$ with $\beta\cdot  \bar z$ in $\Delta_H(\bar z)$
gives a vanishing contribution due to  (\ref{inv-rel1}).
There are other contractions at the same order, but
they cancel between the numerator and the denominator.}

This computation yields
\ba
&&
\f{\langle W(R)\cdot \Ocal_\Delta\rangle_{G,\tau}}{
 \langle W(R)\rangle_{G,\tau}}
 =
\left(\f{g}{8\pi} \right)^\Delta
\Bigg[
P_\Delta(w)+\f{2}{g^2}
\mathop{\sum_{\alpha>0}}_{\langle \alpha,w\rangle\neq 0}
\f{\langle \alpha,\alpha\rangle}{\langle\alpha,w\rangle}
\hat\alpha\cdot
\p P_\Delta(w)
\Bigg].
\label{matrix-result}
\ea
This is the final result to next to leading order in  the strong coupling expansion of the correlator  of the circular Wilson loop $W(R)$
with an arbitrary chiral primary $\Ocal_\Delta$ in ${\cal N}=4$ super Yang-Mills with gauge group $G$. 

By using the formula (\ref{wardiden}), we find that the scaling weight of a circular Wilson loop $W(R)$ at strong coupling is given by
\ba
h_W(R,\tau)=-\f{g^2}{48\pi^2}
\Bigg[ \langle {w},{w}\rangle 
 +\f{8}{g^2}\dim(G/H)\Bigg]\,,
 \label{scalewilson}
\ea
where $w$ is the highest weight in the representation $R$.

\section{\texorpdfstring{$S$}{S}-duality of correlators}
\label{sec-compare}

In this section we demonstrate that the computations we have performed   for 't Hooft and Wilson loop correlators in the previous sections exactly map to each other under the conjectured action of $S$-duality. These results exhibit  $S$-duality in ${\cal N}=4$ super Yang-Mills with arbitrary gauge group $G$ on correlation functions, and extends the recent results in  \cite{Gomis:2009ir}, which demonstrated that the expectation value of a circular 't Hooft operator and a circular Wilson operator are exchanged under electric-magnetic duality.

 In section \ref{sec-thooft}, the correlator
 of a circular  't Hooft loop operator $T(\LR)$ and a chiral primary operator  $\Ocal_\Delta=g^{-\Delta}P_\Delta(Z)$
 was calculated to next to leading order in the weak coupling expansion, yielding  the result (\ref{thooft-parametrizationa3}),
which we reproduce here:
\ba
{\langle T(\LR) \cdot \Ocal_\Delta\rangle_{G,\tau}\over \langle T(\LR) \rangle_{G,\tau}}=
\left({g|\tau|\over 8\pi}\right)^\Delta\left[ P_\Delta(B)+\f{2}{g^2|\tau|^2}
\mathop{\sum_{\alpha>0}}_{\alpha(B)\neq 0}
\f{\langle\alpha,\alpha\rangle}{\alpha(B)} \hat\alpha\cdot \p
 P_\Delta(B)\right]\,.
\label{thooft-result}
\ea

In order to demonstrate $S$-duality, we need the result of the   correlator  for the 
dual operators in the theory with gauge group $\LG$ and coupling constant $\Ltau$. Using the computation in (\ref{matrix-result}), 
we find that the strong coupling expansion of the correlator of a circular Wilson loop operator
$W(\LR)$
and the chiral primary operator $\LL \Ocal_\Delta\equiv (\Lgc)^{-\Delta}\cdot
\LL P_\Delta(\LL Z)$
  is given by
\ba
&&\f{\langle W(\LR)\cdot\LL \Ocal_\Delta\rangle_{\LG,\LL \tau}}{
 \langle W(\LR)\rangle_{\LG,\LL\tau}}
\nn\\
&&
~~~~~~ =
\left(\f{\Lgc}{8\pi} \right)^\Delta\hskip-2pt
\left[
\LL P_\Delta(\Lw)+\f{2}{(\Lgc)^2}\hskip-4pt
\mathop{\sum_{\Lalpha>0}}_{\langle \Lalpha,\Lw\rangle\neq 0}\hskip-9pt
\f{\langle\Lalpha,\Lalpha\rangle}{\langle\Lalpha,\Lw\rangle}
\LL \hat\alpha\cdot
\p\hskip+1pt\LL P_\Delta(\Lw)
\right].
\label{wilson-result-L}
\ea
We recall that under $S$-duality
\ba
\LLtau=-\f{1}{n_\gfrak \tau}\,, \qquad \Longrightarrow (\Lgc)^2=n_\gfrak\, g^2|\tau|^2\,,
\label{transform1}
\ea 
and the gauge groups $G$ and $\LG$ are exchanged. 
Also, as discussed in section \ref{sec-OPE}, $S$-duality induces the following transformations
\ba
\LL P_\Delta(\Lw)&=&
n_{\gfrak}^{-\Delta/2}
P_\Delta(B),\nonumber\\
\Lw&=&
n_\gfrak^{-1/2}
\Rcal(B), \label{transform2}\\
\Lalpha&=&n_\gfrak^{-1/2}\Rcal(\hat \alpha)\,,
\nonumber
\ea
where $\hat \alpha\equiv 2\alpha/|\alpha|^2$ is the coroot
corresponding to $\alpha$ and $\Rcal$ is the linear transformation defined in (\ref{lineartransform}).  
The two expressions in (\ref{thooft-result}) and (\ref{wilson-result-L}) map into each other under the transformations
(\ref{transform1}) and (\ref{transform2}).

In  \cite{Gomis:2009ir},
the prediction of $S$-duality for the expectation values of
loop operators
\ba 
\langle T(\LR)\rangle_{G,\tau}= \langle
W(\LR)\rangle_{\LG,\Ltau}\,,
\ea
was demonstrated
to next to leading order in the coupling constant expansion.
By combining this with the above agreement,
we conclude that  the 't Hooft and Wilson loop correlation functions transform as predicted by $S$-duality
\ba
\langle T(\LR) \cdot \Ocal_\Delta\rangle_{G,\tau}
=
\langle W(\LR)\cdot\LL \Ocal_\Delta\rangle_{\LG,\LL \tau} \,.
\label{finales}
\ea
We have explicitly exhibited this to next to leading order in the coupling constant expansion. Furthermore, this   implies
that  the semiclassical scaling weight of the 't Hooft operator (\ref{scaling-result})
exactly reproduces the scaling weight of the dual Wilson operator  evaluated at strong coupling (\ref{scalewilson}) under the action of $S$-duality:
\ba
h_T(\LR,\tau)=h_W(\LR,\Ltau).
\ea
These are the main results of this paper.

Finally let us discuss the OPE coefficients.
In Appendix \ref{2point} we show that
the two and three-point functions of chiral primary operators
are invariant under $S$-duality.
Since the OPE coefficients and  the correlators of a loop operator  with a chiral primary operator  
are related by the matrix of two-point functions of the local operators, our
results   imply that the OPE coefficients (\ref{OPEloopy})
also match up to the next-to-leading order under the $S$-duality 
transformation:
\ba
b_\Delta(\LR,\tau)=\LL c_\Delta(\LR,\Ltau).
\ea

We have thus found the precise matching  under $S$-duality
of a number of physical observables involving circular 't Hooft and Wilson loop operators.
This provides a quantitative demonstration of the action of electric-magnetic duality on   correlation functions in ${\cal N}=4$ super Yang-Mills
with an arbitrary gauge group $G$.

\subsection*{Acknowledgments}

We are very grateful to Diego Trancanelli for collaboration
at the initial stage of the project.
We also thank Anton Kapustin for useful discussions.
Research at Perimeter Institute is
supported in part by the Government of Canada through NSERC and by
the Province of Ontario through MRI. J.G. also acknowledges further
support from an NSERC Discovery Grant. 
\vfill\eject


\appendix

\section{Weyl transformation between metrics}
\label{sec-conformalfactor}

In this Appendix we discuss the   Weyl transformation 
relating $\R^4$ and $AdS_2\times S^2$.

Let us parametrize $\R^4$
using two sets of polar coordinates so that
\be
ds^2_{\mathbb{R}^4}=dr^2+r^2d\psi^2+dx^2+x^2 d\phi^2. 
\ee
These coordinates are relevant for a circular loop, which we take to be located at
  $r=a$ and $x=0$. 
By making the following change of coordinates
\ba
\begin{array}{c}
\tilde r^2=\displaystyle
{(r^2+x^2-a^2)^2+4a^2x^2\over 4a^2}
=
\displaystyle
{a^2\over(\cosh\rho-\cos\theta)^2}\,,\qquad
\\
\\
r=\tilde r\sinh\rho\,,\qquad
x=\tilde r\sin\theta\,,
\end{array}
\ea
we find the metric
\be 
ds^2_{\mathbb{R}^4}=\tilde r^2\left(d\rho^2+\sinh^2\rho\,d\psi^2+
d\theta^2+\sin^2\theta\,d\phi^2\right)\,,
\label{conff}
\ee
which is conformal to  $AdS_2\times S^2$ in global coordinates. 
Note that the loop, which was located at $r=a,\, x=0$ in $\R^4$, gets mapped to the conformal boundary of $AdS_2\times S^2$, namely the boundary of the Poincar\'e disk.

In the absence of conformal anomaly,
a dimension $\Delta$ 
scalar operator ${\cal O}_\Delta$ transforms as  ${\cal O}_\Delta\rightarrow \tilde{r}^{-J} {\cal O}_\Delta$
under the Weyl transformation (\ref{conff}). 
This proves the position dependence (\ref{correlacircleold}) of the correlator
on $\R^4$. The same Weyl transformation can be used to write down  the form of the correlator of the loop operator with the stress-energy tensor  on $\R^4$   from the $AdS_2\times S^2$ correlator (\ref{scaleweight}).

\section{Chiral primary operators and \texorpdfstring{$S$}{S}-duality}
\label{generators}

Chiral primary operators and their $S$-duality transformation
in $\Ncal=4$ super Yang-Mills with gauge group $G$
play a central role in the current work.  In this Appendix
we supplement  the minimal amount of information
given in section \ref{sec-OPE} with more details and examples.
 
Let us consider the subspace of the Coulomb branch
where only the combination of scalar fields $Z=\phi_1+i\phi_2$ is excited.
The gauge group $G$ is generically broken to $U(1)^r$,
and $Z$ takes expectation values in the Cartan subalgebra $\tfrak_\C$,
and are identified by the action of the Weyl group.

The massless fields $\varphi^i$ relevant to us are the
components of $Z$ in the Cartan subalgebra directions.
Let us canonically normalize them by expanding $Z$ as
$Z=g \varphi^i H_i$ so that the kinetic term in the Lagrangian reads
\ba
\Lcal= |\p_\mu \varphi^i|^2+\ldots.
\ea
Since the low-energy physics is that of an abelian theory
with gauge group $U(1)^r$, $S$-duality acts as
ordinary electric-magnetic duality.
To see how this works let us consider the dual theory with dual gauge group $\LG$.
If we expand the dual scalar as $\LL Z=\Lgc\, \LL \varphi^i\, \LL H_i$,
the kinetic term is
\ba
\LL \Lcal= |\p_\mu \LL \varphi^i|^2+\ldots.
\ea
We identify $\LL\varphi$ with $\varphi$ via
\ba
\LL\varphi=\Rcal \varphi
\ea
using the linear transformation introduced below (\ref{lineartransform}).
The map $\Rcal$ is norm-preserving as necessary for the invariance
of the kinetic term, and the choice of $\Rcal$ is unique up to the Weyl
group action.

The gauge invariant coordinates of the moduli space for the
original gauge theory are provided
by the $r$ generators $P_i$ of the invariant polynomial ring
(\ref{ring-gen}).
They should be identified with the gauge invariant coordinates
in the dual theory according to
\ba
\LL P_i(\LL\varphi)=P_i(\varphi).
\ea
In terms of the scalar $Z$ whose normalization is
such that it has a kinetic term $g^{-2} \tr (\p_\mu\bar Z \p^\mu Z)$,
and its counterpart for $\LL Z$ in the dual theory, 
the $S$-duality map of the chiral primaries is given by
\ba
\oo{(\Lgc)^{\nu_i}}
\LL P_i(\LL Z)\longleftrightarrow 
\oo{g^{\nu_i}}P_i(Z).
\ea
This explains the coupling dependence in (\ref{Sact}).

In the following we illustrate our considerations
by explicitly writing down chiral primary operators
for several choices of gauge group.
Note that $G$-invariant polynomials on $\gfrak_\C$
and  Weyl-invariant polynomials on $\tfrak_\C$
are in one-to-one correspondence.
For exceptional groups
it is more convenient to use the latter description,
and this is what we   do below.
\bi
\item $G=SU(n)$.

In this case the generators of the chiral ring are simply single trace operators
\ba
P_i(Z)=\tr\, Z^{i+1},~~~~i=1,2,\ldots,n-1
\ea
with $\nu_i=i+1$.

\item $G=SO(2n+1)$ and $G=Sp(n)$.

For these groups, the trace of an odd power of the matrix $Z$ vanishes.
Thus the generators are given by the trace of the even powers of $Z$:
\ba
P_i(Z)=\tr\, Z^{2i},~~i=1,2,\ldots, n.
\ea
Their conformal dimensions are given by $\nu_i=2i$.
The Lie algebras of the two gauge groups are exchanged under $S$-duality.

\item $G=SO(2n)$.

For the even orthogonal group, in addition to the trace
of an even power of $Z$'s one can consider the Pfaffian.
The generators are
\ba
P_i(Z)=\tr\, Z^{2i},&&~~i=1,2,\ldots, n-1,\\
P_n(Z)={\rm Pf}(Z)&\equiv&\f{1}{2^n n!} \epsilon^{i_1i_2\ldots i_{2n-1}i_{2n}}
 Z_{i_1 i_2}\ldots Z_{i_{2n-1}i_{2n}}.
\ea
These have conformal dimensions $\nu_i=2i$ for $i=1,\ldots, n-1$,
and $\nu_n=n$.

\item $G=G_2$.

Here we choose to be less explicit
and describe chiral primary operators in terms of
Weyl invariant polynomials on $\tfrak$.
The Cartan subalgebra $\tfrak$ is two-dimensional
and can be identified with the plane $x_1+x_2+x_3=0$
in $\R^3$.
The Weyl group is generated by the permutations of the $x_i$'s
and the overall sign change.
Thus as generators of the Weyl-invariant polynomials on $\tfrak$,
we  can take \cite{Argyres:2006qr}
\ba
P_1=x_1^2+x_2^2+x_3^2,~~~~P_2=x_1^2 x_2^2 x_3^2
\ea
with $\nu_1=2, \nu_2=6$.
According to (\ref{P-LP}), under $S$-duality they transform to
\ba
\LL P_1=P_1,~~\LL P_2=-P_2+\oo{54} P_1^3
\ea
since $\Rcal$ acts as
$(x_1,x_2,x_3)\mapsto 3^{-1/2}(x_2-x_3, x_1-x_2,x_3-x_1)$
\cite{Argyres:2006qr}.

\ei

\section{Complex and normal matrix models
for  any \texorpdfstring{$G$}{G}}\label{app-normal}

The aim of this Appendix is to derive the relation
between the complex and normal matrix models for general $G$,
as used in (\ref{complex-normal2}).
This is done by generalizing the derivation of the
relation in the $U(n)$ case given in \cite{Okuyama:2006jc}.

First we decompose the complex variable $z$ into the real and imaginary parts:
\ba
z=x+iy\in \gfrak_\C,~~x,y\in \gfrak.
\ea
Then the complex matrix model integral is defined by
\ba
I_P= \int [dx][ dy]e^{-\f{2}{g^2}(\tr\, x^2+\tr\, y^2)
}
\Tr_R e^x
P(x+iy),
\ea
where $P$ is an arbitrary invariant polynomial on $\gfrak_\C$.
Let us introduce an orthonormal basis $T_a$ of $\gfrak$ 
satisfying
\ba
\tr( T_a T_b)= \delta_{ab}
\ea
and write
\ba
x=x^a T_a,~~~y=y^a T_a.
\ea
The measure is then
\ba
[dx][dy]=\prod_a dx^a dy^a.
\ea
By writing
\ba
P(x+iy)=e^{iy^a\f{\p}{\p x^a}}P(x),
\ea
we can integrate out $y$ so that the integral is now
\ba
I_P
=
\left( \f{\pi g^2}{2}\right)^{\dim G/2}
\int [dx]
 e^{-\f{2}{g^2} \tr\, x^2
}
\Tr_R e^{x} e^{-\f{g^2}8 \nabla_\gfrak^2}
P(x).
\ea
Here $\nabla_\gfrak^2$ is the Laplacian on $\gfrak$.
To further reduce the integral,
let us represent $\gfrak$ as a fibration of 
$G/T$
over $\mathfrak{t}$ (the fibration degenerates on a set of measure
zero):
\ba
x=\sfg^\mo\lambda \sfg,~~~\sfg\in G\,,
\ea
where $T$ is the maximal torus of $G$.
Here $\lambda\in \mathfrak t$, and $g$ parametrizes the fiber,
which is the adjoint orbit of $\lambda$.
If we expand  $\sfg^\mo d\sfg$
as
\ba
\sfg^\mo d\sfg=i(d\xi^i H_i+ d\xi^\alpha H_\alpha)
\ea
in the Cartan basis,
the metric is then
\ba
 ds^2_{\gfrak}
=ds^2_{\mathfrak{t}}
+2\sum_{\alpha>0}
\alpha(\lambda)^2
\tr(E_\alpha E_{-\alpha})
 d\xi^\alpha d\xi^{-\alpha}.
\ea
Let us normalize $H_i$ so that $\langle H_i, H_j\rangle=\delta_{ij}$.
We can write the Laplacian on $\gfrak$ as
\ba
\nabla^2_\gfrak=\f{1}{\Delta(\lambda)^2}  \f{\p}{\p \lambda^i}\Delta(\lambda)^2
\f{\p}{\p \lambda^i}+(\hbox{derivatives in $G/T$-directions}),
\ea
where
\ba
\Delta(\lambda)=\prod_{\alpha>0}\alpha(\lambda).
\ea
Note that $\Delta(\lambda)$ is skew-symmetric with respect to
the Weyl group.
In fact any skew-symmetric polynomial has to be
divisible by $\Delta(\lambda)$ because such a polynomial
vanishes along the hyperplane $\alpha(\lambda)=0$
fixed by the Weyl reflection associated with $\alpha$.
Since the metric on $\mathfrak t$ is Weyl invariant, the polynomial
\ba
\sum \f{\p}{\p \lambda^i}\f{\p}{\p \lambda^i}\Delta(\lambda)
=\nabla^2_{\mathfrak t}\Delta(\lambda)
\ea
is also skew-symmetric.
The polynomial however has a lower degree than $\Delta$,
so it has to vanish, i.e., $\Delta$ is harmonic on
 $\mathfrak t$ \cite{1980InMat..56...93M}.
Using the fact that $\Delta$ is harmonic, we can write
\ba
\nabla^2_\gfrak=\f{1}{\Delta(\lambda)}  
\nabla^2_{\mathfrak t}
\Delta(\lambda)+(\hbox{derivatives in $G/T$-directions}).
\ea
Also note that
the quotient metric on $G/T$ is given by
\ba
ds^2_{G/T}=2\sum_{\alpha>0}
\tr(E_\alpha E_{-\alpha})
d\xi^\alpha d\xi^{-\alpha}.
\label{G/T}
\ea
Hence the volume form on the orbit of $\lambda$ is given by
\ba
\Delta(\lambda)^2{\rm vol}(G/T),
\ea
where ${\rm vol}(G/T)$ is the volume form on $G/T$
constructed from the metric (\ref{G/T}).
Thus
\ba
I_P&=&
\left( \f{\pi g^2}{2}\right)^{\dim G/2}
\f{{\rm Vol}(G/T)}{|\Wcal|}
\int_{\mathfrak t} [d\lambda]\Delta(\lambda)^2
 e^{-\f{2}{g^2} \langle \lambda,\lambda\rangle}
\Tr_R e^\lambda 
\Delta(\lambda)^\mo
e^{-\f{g^2}8 \nabla^2_{\mathfrak t}}
\Delta(\lambda)P(\lambda),
\nn
\ea
where $|\Wcal|$ is the order of the Weyl group $\Wcal$,
and $[d\lambda]=\prod_i d\lambda^i$.
In order to keep the equations simple,
from now on we will neglect those prefactors which cancel in
(\ref{complex-normal2}).
Let us define $\eta=\f{\sqrt 2}g\lambda$.
Then
\ba
I_P\propto \int_{\mathfrak t}[d\eta] \Delta(\eta)
e^{-\langle \eta,\eta\rangle}
\Tr_R e^{\f{g}{\sqrt 2}\eta} e^{-\oo 4\nabla_{\mathfrak t}^2}
\Delta(\eta)P\left(\f{g}{\sqrt 2}\eta\right).
\ea
Using the identity
\ba
e^{-\oo 4 \p_\eta^2} f(\eta)=e^{\half \eta^2}
f\left(\f{\eta-\p_\eta}2\right)e^{-\half \eta^2} \label{okuyama-identity}
\ea
that holds for any function $f(\eta)$,
we get
\ba
I_P&\propto&
\int_{\mathfrak t} [d\eta]\Delta(\eta)
 e^{-\f{1}{ 2} \langle \eta,\eta\rangle}
\Tr_R e^{\f{g}{\sqrt 2}\eta}
\nn\\
&&~~~~\times
P\left(\f{g}{\sqrt 2}\left(\f{\eta^i}2-\f{1}2\f{\p}{\p \eta^i}\right)\right)
\Delta\left(\f{\eta^i}2-\f{1}2\f{\p}{\p \eta^i}\right)
 e^{-\f{1}{ 2} \langle \eta,\eta\rangle}.
\ea
Using (\ref{okuyama-identity}) and harmonicity,
we can write
\ba
\Delta\left(\f{\eta^i}2-\f{1}2\f{\p}{\p \eta^i}\right)
 e^{-\f{1}{2} \langle \eta,\eta\rangle}
=\Delta(\eta)
 e^{-\f{1}{ 2} \langle \eta,\eta\rangle}:=\Psi(\eta).
\ea
Then the integral is now
\ba
I_P&\propto&
\int_{\mathfrak t} [d\eta]
\Psi(\eta)
\Tr_R e^{\f{g}{\sqrt 2}\eta}
P\left(\f{g}{\sqrt 2}\left(\f{\eta^i}2-\f{1}2\f{\p}{\p \eta^i}\right)\right)
\Psi(\eta).
\ea
The differential operator can be interpreted as creation
operators in an oscillator system
\ba
[a_i, a^\dagger_j]=\delta_{ij}.
\ea
Thus $\Psi(\eta)$ is the wave function for the state
$|\Psi\rangle\propto \Delta(a^\dagger)|0\rangle$.
The integral now takes the form
\ba
I_P&\propto& \langle \Psi| \Tr_R e^{\f{g}{ 2}(a+a^\dagger)}
P\left(\f{g}{ 2}a^\dagger\right)
|\Psi\rangle
\nn\\
&=&\sum_v n(v) \langle \Psi|
e^{-\f{g^2}8 \langle v,v\rangle}
 e^{\f{g}2 v(a)}
e^{\f{g}2 v(a^\dagger)}
 P\left(\f{g}2 a^\dagger
\right) |\Psi\rangle,
\ea
where in the second line we wrote the character
as a sum over weights $v$ with multiplicity $n(v)$.
By using the completeness of coherent states
\ba
1\propto\int \prod_i d^2 z^i |z\rangle \langle z|,~~~
a^i |z\rangle=z^i |z\rangle,
\ea
we can write the integral as
\ba
I_P&\propto&
\sum_{{v}} n(v) e^{-\f{g^2\langle {v},{v}\rangle}8}
\int_{{\mathfrak t}_\C} [dz]
\Delta(z)\Delta(\bar z)
e^{-\langle \bar z,z\rangle}
e^{\f{g}{ 2} {v}(z+\bar z)}
P\left(\f{g}{ 2} z\right). \label{normal-1}
\ea

Since we are interested in the strong coupling limit,
we further transform  the normal matrix model
into a form where the strong coupling expansion is
easy to perform by shifting $z\ra z+\half g {v}$ in (\ref{normal-1}):
\ba
I_P\propto 
\sum_{v} n(v) e^{\f{g^2}8 \langle {v},{v}\rangle}
\int_{{\mathfrak t}_\C} [dz]
\left|\Delta\left( {v}+\f{2}g z\right)\right|^2
e^{-\langle \bar z,z\rangle}
P\left(\f{g^2}4{v}+\f{g}{ 2} z\right).
\ea
Here we have identified $\mathfrak t$ with $\mathfrak t^*$ using the metric.
By taking the ratio $I_P/I_1$, we obtain the relation
(\ref{complex-normal2}).

\section{\texorpdfstring{$S$}{S}-duality of 2- and 3-point functions of CPO's}
\label{2point}

In this Appendix we  show that the two and three-point functions of 
chiral primary operators $\Ocal^{(i)}_{\Delta}$(see eqn. 
(\ref{CPOgeneral})) transform according to the $S$-duality conjecture
\ba
\langle \Ocal^{(1)}_{\Delta}\cdot\bar{\Ocal}^{(2)}_{\Delta}\rangle_{G,\tau}&=&
\langle \LL \Ocal^{(1)}_{\Delta}\cdot \bar{\LL\hspace{.3mm} \Ocal}^{(2)}_{\Delta}
\rangle_{\LG,\Ltau}\,,
\label{dual-2pt}\\
\langle \Ocal^{(1)}_{\Delta_1}\cdot\Ocal^{(2)}_{\Delta_2}\cdot
\bar{\Ocal}^{(3)}_{\Delta_1+\Delta_2}\rangle_{G,\tau}&=&\langle \LL 
\Ocal^{(1)}_{\Delta_1}\cdot \LL \Ocal^{(2)}_{\Delta_2}\cdot \bar{\LL
\hspace{.2mm} 
\Ocal}^{(3)}_{\Delta_1+\Delta_2}\rangle_{\LG,\Ltau}\,.
\label{dualidadscalar}
\ea
These correlation functions, which are independent of the coupling 
constant, can be computed using a complex Gaussian matrix model where 
the matrix $z$ takes values in the complexified Lie algebra $\gfrak_\C
$  \cite{Okuyama:2006jc}.

The spatial dependence of the correlator is fixed by conformal 
invariance and the two and three-point correlators are given
respectively by the 
matrix integrals
\ba
\f{
\displaystyle
\int_{\gfrak_\C }[dz] e^{-\tr(\bar z z)}
P^{(1)}(z)
\bar{P^{(2)}(z)}
}{
\displaystyle
\int_{\gfrak_\C}[dz]e^{-\tr(\bar z z)}}
\label{2pt}
\ea
and
\ba
\f{
\displaystyle
\int_{\gfrak_\C }[dz] e^{-\tr(\bar z z)}
P^{(1)}(z)P^{(2)}(z)
\bar{P^{(3)}(z)}
}{
\displaystyle
\int_{\gfrak_\C}[dz]e^{-\tr(\bar z z)}}\,.
\label{3pt}
\ea

Any complex matrix $z\in \gfrak_\C$ can be decomposed as 
\ba
z=\mathsf{g} b\mathsf{g}^{-1}~~~~~~ b\in {\mathfrak b}, ~~~~g\in G\,,
\label{decomposition}
\ea
where $b$ belongs to the Borel subalgebra ${\mathfrak b}
=\tfrak_\C\oplus
(\oplus_{\alpha>0} \gfrak_\alpha)$ of $\gfrak_\C$, and
$\gfrak_\alpha$ is generated by the raising operator $E_{\alpha}$ in 
the Weyl basis. The Borel subalgebra   generalizes the   subgroup of 
upper triangular
matrices in $\mathfrak{u}(n)_\C$ to an arbitrary Lie algebra $\gfrak_\C$.

We recall that the an invariant polynomial $P(z)$
can be written in terms of a rank-$\Delta$ invariant symmetric tensor
on the Lie algebra as
\ba
P(z)=K_{a_1\ldots a_\Delta}z^{a_1}\ldots z^{a_\Delta}
=K(\stackrel{\Delta}{\overbrace{z,\ldots,z}})\,.
\ea
We claim that when $P$ is evaluated on an element of the Borel subalgebra $b
\in{\mathfrak b}$,     $P$ is just a function of the field components
$\lambda
$ 
in the Cartan subalgebra $\tfrak_\C$. This follows from the fact that 
$K_{a_1\ldots a_\Delta}$ is an invariant tensor on $\gfrak$, which 
implies that
\ba
\sum_{l=1}^{\Delta} K(z_1, \ldots, z_{l-1},[z,z_l],z_{l+1},\ldots,z_
\Delta)=0\,.
\ea
If we let $z_1=E_{\alpha_1},\ldots, z_s=E_{\alpha_s}, z_{s+1}=\ldots=z_
\Delta=\lambda, z=\lambda'$, where $\lambda,\lambda'\in \tfrak_\C$
and $E_{\alpha}$ are ladder 
operators in the Cartan basis,  then invariance of $K$ implies
\ba
 [ (\alpha_1+\ldots+\alpha_s)(\lambda')]
K(E_{\alpha_1},\ldots,E_{\alpha_s},\lambda,
\ldots,\lambda)=0\,.
\ea
In the  Borel subalgebra ${\mathfrak b}$  all roots are positive, and   
therefore
  $ (\alpha_1+\ldots+\alpha_s)\neq 0$. This implies that
\ba
K(E_{\alpha_1},\ldots,E_{\alpha_s},\lambda,\ldots,\lambda)=0, ~~~~~~ {\rm for}~ 
s=1,\ldots,\Delta\,.
\ea
This demonstrates that  any invariant polynomial 
evaluated on the Borel subalgebra ${\mathfrak b}$ depends only on 
the field components 
in the Cartan subalgebra $\tfrak_\C$:
\ba
P(b)=P(\lambda),~~~b\in {\mathfrak b},~~\lambda \in \tfrak_{\C},~~b-\lambda
\in \mathop\oplus_{\alpha>0} \gfrak_\alpha.
\ea

By using the decomposition in (\ref{decomposition}) we can compute the 
Jacobian of the change of variables      (see Appendix A.33 in 
\cite{Mehta}) and write the integrals in (\ref{2pt}) and (\ref{3pt})   
completely in terms of integration over the Cartan subalgebra $\tfrak_\C$
\ba
\f{
\displaystyle
\int_{\gfrak_\C }[dz] e^{-\tr(\bar z z)}
P^{(1)}(z)
\bar{P^{(2)}(z)}
}{
\displaystyle
\int_{\gfrak_\C}[dz]e^{-\tr(\bar z z)}}=\f{
\displaystyle
\int_{\tfrak_\C} [dz] |\Delta(z) |^2
e^{-\langle\bar z, z\rangle}
P^{(1)}(z)
\bar{P^{(2)}(z)}
}{
\displaystyle
\int_{\tfrak_\C}[dz]
|\Delta(z)|^2
e^{-\langle \bar z, z\rangle}}\,,
\label{boreloa}
\ea
\ba
&&
\f{
\displaystyle
\int_{\gfrak_\C }[dz] e^{-\tr(\bar z z)}
P^{(1)}(z)P^{(2)}(z)
\bar{P^{(3)}(z)}
}{
\displaystyle
\int_{\gfrak_\C}[dz]e^{-\tr(\bar z z)}}
\nn\\
&&
~~~~~~~~~~~
=
\f{
\displaystyle
\int_{\tfrak_\C} [dz] |\Delta(z) |^2
e^{-\langle \bar z, z\rangle}
P^{(1)}(z)P^{(2)}(z)
\bar{P^{(3)}(z)}
}{
\displaystyle
\int_{\tfrak_\C}[dz]
|\Delta(z)|^2
e^{-\langle \bar z, z\rangle}}\,,
\label{borelo}
\ea
where
\ba
\Delta(z)\equiv \prod_{\alpha>0}\alpha(z)\,.
\ea

We now need to show that these expressions transform  properly  under 
the $S$-duality map
(\ref{P-LP}).
Indeed, if we define
\ba
\LL z=\Rcal z
\ea
for $z\in \tfrak_\C$, then
\ba
[d\LL z]=[dz],~~~\langle \LL \,\bar z, \LL z\rangle=
\langle\bar z, z\rangle,~~
\LL \Delta(\LL z)\equiv \prod_{\Lalpha>0}\Lalpha(\LL z).
\ea
Moreover, we have that
\ba
\LL P^{(i)}(\LL z)=P^{(i)}(z),~~  \LL \Delta(\LL z)= ({\rm prefactor})
\Delta(z).
\ea
The prefactor cancels out between the numerator and denominator in 
(\ref{boreloa}) and (\ref{borelo}).
Thus under $S$-duality we get that
\ba
\left\langle P^{(1)}\left(\oo{g} Z\right) \bar{P^{(2)}\left(\oo g Z\right)}
\right
\rangle_{G,\tau}
=
\left\langle \LL P^{(1)}\left(\oo \Lgc  \LL Z\right)
    \bar{\LL P^{(2)}\left(\oo{\Lgc} \LL Z\right)} \right\rangle_{\LG,\Ltau}
\ea
and
\ba
&&\left\langle P^{(1)}\left(\oo{g} Z\right) P^{(2)}\left(\oo{g} Z\right) 
\bar{P^{(3)}\left(\oo g Z\right)}\right
\rangle_{G,\tau}
\nn\\
&&~~~~~~~~~~=
\left\langle \LL P^{(1)}\left(\oo \Lgc  \LL Z\right)
\LL P^{(2)}\left(\oo \Lgc  \LL Z\right)  \bar{\LL P^{(3)}\left(\oo{\Lgc} \LL Z
\right)} \right\rangle_{\LG,\Ltau}.
\ea
This implies that the two and three-point functions of chiral primary 
operators in ${\cal N}=4$ super Yang-Mills transform 
according to (\ref{dual-2pt}) and (\ref{dualidadscalar}) under $S$-duality 
as conjectured.
\vfill\eject

\bibliography{draft}

\providecommand{\href}[2]{#2}\begingroup\raggedright\begin{thebibliography}{10}

\bibitem{Wilson:1974sk}
K.~G. Wilson, ``{Confinement of Quarks},''
\href{http://dx.doi.org/10.1103/PhysRevD.10.2445}{{\em Phys. Rev.} {\bf D10}
  (1974)  2445--2459}.

\bibitem{'tHooft:1977hy}
G.~'t~Hooft, ``{On the Phase Transition Towards Permanent Quark Confinement},''
\href{http://dx.doi.org/10.1016/0550-3213(78)90153-0}{{\em Nucl. Phys.} {\bf
  B138} (1978)  1}.

\bibitem{Montonen:1977sn}
C.~Montonen and D.~I. Olive, ``{Magnetic Monopoles as Gauge Particles?},''
\href{http://dx.doi.org/10.1016/0370-2693(77)90076-4}{{\em Phys. Lett.} {\bf
  B72} (1977)  117}.

\bibitem{Witten:1978ma}
E.~Witten and D.~I. Olive, ``{Supersymmetry Algebras That Include Topological
  Charges},''
\href{http://dx.doi.org/10.1016/0370-2693(78)90357-X}{{\em Phys. Lett.} {\bf
  B78} (1978)  97}.

\bibitem{Osborn:1979tq}
H.~Osborn, ``{Topological Charges for N=4 Supersymmetric Gauge Theories and
  Monopoles of Spin 1},''
\href{http://dx.doi.org/10.1016/0370-2693(79)91118-3}{{\em Phys. Lett.} {\bf
  B83} (1979)  321}.

\bibitem{Goddard:1976qe}
P.~Goddard, J.~Nuyts, and D.~I. Olive, ``{Gauge Theories and Magnetic
  Charge},''
\href{http://dx.doi.org/10.1016/0550-3213(77)90221-8}{{\em Nucl. Phys.} {\bf
  B125} (1977)  1}.

\bibitem{Kapustin:2005py}
A.~Kapustin, ``{Wilson-'t Hooft operators in four-dimensional gauge theories
  and S-duality},'' \href{http://dx.doi.org/10.1103/PhysRevD.74.025005}{{\em
  Phys. Rev.} {\bf D74} (2006)  025005},
\href{http://arxiv.org/abs/hep-th/0501015}{{\tt arXiv:hep-th/0501015}}.

\bibitem{Gomis:2009ir}
J.~Gomis, T.~Okuda, and D.~Trancanelli, ``{Quantum 't Hooft operators and
  S-duality in N=4 super Yang-Mills},''
\href{http://arxiv.org/abs/0904.4486}{{\tt arXiv:0904.4486 [hep-th]}}.

\bibitem{Intriligator:1998ig}
K.~A. Intriligator, ``{Bonus symmetries of N = 4 super-Yang-Mills correlation
  functions via AdS duality},''
  \href{http://dx.doi.org/10.1016/S0550-3213(99)00242-4}{{\em Nucl. Phys.} {\bf
  B551} (1999)  575--600},
\href{http://arxiv.org/abs/hep-th/9811047}{{\tt arXiv:hep-th/9811047}}.

\bibitem{Argyres:2006qr}
P.~C. Argyres, A.~Kapustin, and N.~Seiberg, ``{On S-duality for
  non-simply-laced gauge groups},'' {\em JHEP} {\bf 06} (2006)  043,
\href{http://arxiv.org/abs/hep-th/0603048}{{\tt arXiv:hep-th/0603048}}.

\bibitem{Shifman:1980ui}
M.~A. Shifman, ``{Wilson Loop in Vacuum Fields},''
\href{http://dx.doi.org/10.1016/0550-3213(80)90440-X}{{\em Nucl. Phys.} {\bf
  B173} (1980)  13}.

\bibitem{Berenstein:1998ij}
D.~E. Berenstein, R.~Corrado, W.~Fischler, and J.~M. Maldacena, ``{The operator
  product expansion for Wilson loops and surfaces in the large N limit},''
  \href{http://dx.doi.org/10.1103/PhysRevD.59.105023}{{\em Phys. Rev.} {\bf
  D59} (1999)  105023},
\href{http://arxiv.org/abs/hep-th/9809188}{{\tt arXiv:hep-th/9809188}}.

\bibitem{Gomis:2008qa}
J.~Gomis, S.~Matsuura, T.~Okuda, and D.~Trancanelli, ``{Wilson loop correlators
  at strong coupling: from matrices to bubbling geometries},''
  \href{http://dx.doi.org/10.1088/1126-6708/2008/08/068}{{\em JHEP} {\bf 08}
  (2008)  068},
\href{http://arxiv.org/abs/0807.3330}{{\tt arXiv:0807.3330 [hep-th]}}.

\bibitem{Donagi:2006cr}
R.~Donagi and T.~Pantev, ``{Langlands duality for Hitchin systems},''
\href{http://arxiv.org/abs/math/0604617}{{\tt arXiv:math/0604617}}.

\bibitem{Gukov:2006jk}
S.~Gukov and E.~Witten, ``{Gauge theory, ramification, and the geometric
  langlands program},''
\href{http://arxiv.org/abs/hep-th/0612073}{{\tt arXiv:hep-th/0612073}}.

\bibitem{Intriligator:1999ff}
K.~A. Intriligator and W.~Skiba, ``{Bonus symmetry and the operator product
  expansion of N = 4 super-Yang-Mills},''
  \href{http://dx.doi.org/10.1016/S0550-3213(99)00430-7}{{\em Nucl. Phys.} {\bf
  B559} (1999)  165--183},
\href{http://arxiv.org/abs/hep-th/9905020}{{\tt arXiv:hep-th/9905020}}.

\bibitem{Witten:1979ey}
E.~Witten, ``{Dyons of Charge e theta/2 pi},''
\href{http://dx.doi.org/10.1016/0370-2693(79)90838-4}{{\em Phys. Lett.} {\bf
  B86} (1979)  283--287}.

\bibitem{Kapustin:2007wm}
A.~Kapustin and N.~Saulina, ``{The algebra of Wilson-'t Hooft operators},''
  \href{http://dx.doi.org/10.1016/j.nuclphysb.2009.02.004}{{\em Nucl. Phys.}
  {\bf B814} (2009)  327--365},
\href{http://arxiv.org/abs/0710.2097}{{\tt arXiv:0710.2097 [hep-th]}}.

\bibitem{Sen:2008yk}
A.~Sen, ``{Entropy Function and AdS(2)/CFT(1) Correspondence},''
  \href{http://dx.doi.org/10.1088/1126-6708/2008/11/075}{{\em JHEP} {\bf 11}
  (2008)  075},
\href{http://arxiv.org/abs/0805.0095}{{\tt arXiv:0805.0095 [hep-th]}}.

\bibitem{Gupta:2008ki}
R.~K. Gupta and A.~Sen, ``{Ads(3)/CFT(2) to Ads(2)/CFT(1)},''
  \href{http://dx.doi.org/10.1088/1126-6708/2009/04/034}{{\em JHEP} {\bf 04}
  (2009)  034},
\href{http://arxiv.org/abs/0806.0053}{{\tt arXiv:0806.0053 [hep-th]}}.

\bibitem{Sen:2008vm}
A.~Sen, ``{Quantum Entropy Function from AdS(2)/CFT(1) Correspondence},''
\href{http://arxiv.org/abs/0809.3304}{{\tt arXiv:0809.3304 [hep-th]}}.

\bibitem{Sen:2009vz}
A.~Sen, ``{Arithmetic of Quantum Entropy Function},''
\href{http://arxiv.org/abs/0903.1477}{{\tt arXiv:0903.1477 [hep-th]}}.

\bibitem{Banerjee:2009af}
N.~Banerjee, S.~Banerjee, R.~Gupta, I.~Mandal, and A.~Sen, ``{Supersymmetry,
  Localization and Quantum Entropy Function},''
\href{http://arxiv.org/abs/0905.2686}{{\tt arXiv:0905.2686 [hep-th]}}.

\bibitem{Gomis:2007fi}
J.~Gomis and S.~Matsuura, ``{Bubbling surface operators and S-duality},'' {\em
  JHEP} {\bf 06} (2007)  025,
\href{http://arxiv.org/abs/0704.1657}{{\tt arXiv:0704.1657 [hep-th]}}.

\bibitem{Drukker:2008wr}
N.~Drukker, J.~Gomis, and S.~Matsuura, ``{Probing N=4 SYM With Surface
  Operators},'' \href{http://dx.doi.org/10.1088/1126-6708/2008/10/048}{{\em
  JHEP} {\bf 10} (2008)  048},
\href{http://arxiv.org/abs/0805.4199}{{\tt arXiv:0805.4199 [hep-th]}}.

\bibitem{Rey:1998ik}
S.-J. Rey and J.-T. Yee, ``{Macroscopic strings as heavy quarks in large N
  gauge theory and anti-de Sitter supergravity},''
  \href{http://dx.doi.org/10.1007/s100520100799}{{\em Eur. Phys. J.} {\bf C22}
  (2001)  379--394},
\href{http://arxiv.org/abs/hep-th/9803001}{{\tt arXiv:hep-th/9803001}}.

\bibitem{Maldacena:1998im}
J.~M. Maldacena, ``{Wilson loops in large N field theories},''
  \href{http://dx.doi.org/10.1103/PhysRevLett.80.4859}{{\em Phys. Rev. Lett.}
  {\bf 80} (1998)  4859--4862},
\href{http://arxiv.org/abs/hep-th/9803002}{{\tt arXiv:hep-th/9803002}}.

\bibitem{Erickson:2000af}
J.~K. Erickson, G.~W. Semenoff, and K.~Zarembo, ``{Wilson loops in N = 4
  supersymmetric Yang-Mills theory},''
  \href{http://dx.doi.org/10.1016/S0550-3213(00)00300-X}{{\em Nucl. Phys.} {\bf
  B582} (2000)  155--175},
\href{http://arxiv.org/abs/hep-th/0003055}{{\tt arXiv:hep-th/0003055}}.

\bibitem{Drukker:2000rr}
N.~Drukker and D.~J. Gross, ``{An exact prediction of N = 4 SUSYM theory for
  string theory},'' \href{http://dx.doi.org/10.1063/1.1372177}{{\em J. Math.
  Phys.} {\bf 42} (2001)  2896--2914},
\href{http://arxiv.org/abs/hep-th/0010274}{{\tt arXiv:hep-th/0010274}}.

\bibitem{Pestun:2007rz}
V.~Pestun, ``{Localization of gauge theory on a four-sphere and supersymmetric
  Wilson loops},''
\href{http://arxiv.org/abs/0712.2824}{{\tt arXiv:0712.2824 [hep-th]}}.

\bibitem{Semenoff:2001xp}
G.~W. Semenoff and K.~Zarembo, ``{More exact predictions of SUSYM for string
  theory},'' \href{http://dx.doi.org/10.1016/S0550-3213(01)00455-2}{{\em Nucl.
  Phys.} {\bf B616} (2001)  34--46},
\href{http://arxiv.org/abs/hep-th/0106015}{{\tt arXiv:hep-th/0106015}}.

\bibitem{Okuyama:2006jc}
K.~Okuyama and G.~W. Semenoff, ``{Wilson loops in N = 4 SYM and fermion
  droplets},'' {\em JHEP} {\bf 06} (2006)  057,
\href{http://arxiv.org/abs/hep-th/0604209}{{\tt arXiv:hep-th/0604209}}.

\bibitem{Giombi:2006de}
S.~Giombi, R.~Ricci, and D.~Trancanelli, ``{Operator product expansion of
  higher rank Wilson loops from D-branes and matrix models},'' {\em JHEP} {\bf
  10} (2006)  045,
\href{http://arxiv.org/abs/hep-th/0608077}{{\tt arXiv:hep-th/0608077}}.

\bibitem{Semenoff:2006am}
G.~W. Semenoff and D.~Young, ``{Exact 1/4 BPS loop: Chiral primary
  correlator},'' \href{http://dx.doi.org/10.1016/j.physletb.2006.10.047}{{\em
  Phys. Lett.} {\bf B643} (2006)  195--204},
\href{http://arxiv.org/abs/hep-th/0609158}{{\tt arXiv:hep-th/0609158}}.

\bibitem{Pestun:2009nn}
V.~Pestun, ``{Localization of the four-dimensional N=4 SYM to a two- sphere and
  1/8 BPS Wilson loops},''
\href{http://arxiv.org/abs/0906.0638}{{\tt arXiv:0906.0638 [hep-th]}}.

\bibitem{Giombi:2009ds}
S.~Giombi and V.~Pestun, ``{Correlators of local operators and 1/8 BPS Wilson
  loops on $S^2$ from 2d YM and matrix models},''
\href{http://arxiv.org/abs/0906.1572}{{\tt arXiv:0906.1572 [hep-th]}}.

\bibitem{Bassetto:2009rt}
A.~Bassetto {\em et al.}, ``{Correlators of supersymmetric Wilson-loops,
  protected operators and matrix models in N=4 SYM},''
\href{http://arxiv.org/abs/0905.1943}{{\tt arXiv:0905.1943 [hep-th]}}.

\bibitem{1980InMat..56...93M}
I.~G. {MacDonald}, ``{The Volume of a Compact Lie Group.},''
  \href{http://dx.doi.org/10.1007/BF01392542}{{\em Inventiones Mathematicae}
  {\bf 56} (1980)  93--95}.

\bibitem{Mehta}
M.~Mehta, ``{Random Matrices},'' {\em Pure and Applied Mathematics Series;
  Third Edition}  .

\end{thebibliography}\endgroup

\end{document}